\documentclass{jfm}
\usepackage{graphicx}
\usepackage{epstopdf, epsfig}
\usepackage[T1]{fontenc}
\usepackage[latin9]{inputenc}
\usepackage{amsmath}
\usepackage{graphicx}
\usepackage{multirow}
\usepackage{xcolor}

\shorttitle{turbulent closure for Jupiter}
\shortauthor{E. Woillez and F. Bouchet}

\title{Barotropic theory for the velocity profile of Jupiter turbulent jets: an example for an exact turbulent closure}

\author{E. Woillez\aff{1}
 \and F. Bouchet \corresp{\email{freddy.bouchet@ens-lyon.fr}}\aff{1}}

\affiliation{\aff{1}Univ Lyon, Ens de Lyon, Univ Claude Bernard, CNRS, Laboratoire de Physique, F-69342 Lyon, France}

\begin{document}

\maketitle

\begin{abstract}
	We model the dynamics of Jupiter's jets by the stochastic barotropic beta-plane model. In this simple framework, by analytic computation of the averaged effect of eddies, we obtain three new explicit results about the equilibrium structure of jets. First we obtain a very simple explicit relation between the Reynolds stresses, the energy injection rate, and the averaged velocity shear. This predicts the averaged velocity profile far from the jet edges (extrema of zonal velocity). Our approach takes advantage of a timescale separation between the inertial dynamics on one hand, and the spin up (or spin down) time on the other hand. Second, a specific asymptotic expansion close to the eastward jet extremum explains the formation of a cusp at the scale of energy injection, characterised by a curvature that is independent from the forcing spectrum. Finally, we derive equations that describe the evolution of the westward tip of the jets. The analysis of these equations is consistent with the previously discussed picture of barotropic adjustment, explaining the relation between the westward jet curvature and the beta effect. Our results give a consistent overall theory of the stationary velocity profile of inertial barotropic zonal jets, in the limit of small scale forcing.

\end{abstract}

\begin{keywords}
Authors should not enter keywords on the manuscript, as these must be chosen by the author during the online submission process and will then be added during the typesetting process (see http://journals.cambridge.org/data/\linebreak[3]relatedlink/jfm-\linebreak[3]keywords.pdf for the full list)
\end{keywords}

\section{Introduction}

The giant gaseous planets like Jupiter and Saturn can be seen as paradigmatic systems to study geostrophic turbulent flows (see
\citep{vasavada2005jovian} for Jupiter). Gallileo and Cassini gave high resolution observations of
Jupiter's troposphere dynamics \citep{salyk2006interaction,porco2003cassini}. The large alternating colored bands at the top of the troposphere are correlated with the zonal wind vorticity. Vortices with a scale of about a thousand kilometers often appear after three dimensional convective activity in the atmosphere. The interaction between those vortices and the zonal jets continuously transfers energy to the barotropic component \citep{ingersoll1981interaction,salyk2006interaction}, and equilibrates the dissipation mechanisms.  The dynamics of large scale jet formation may be qualitatively well understood within the framework of two-dimensional geostrophic turbulence in a $\beta$ plane \citep{PedloskyBook}, although more refined models are needed to understand their quantitative features \citep{li2006interaction,Schneider_Liu_2009JAtS...66..579S}. As the aim of this work is to make progresses in the theoretical understanding of turbulent flows, we consider geostrophic turbulence in the simple barotropic $\beta$ plane model. Despite all its limitations, for instance the lack of dynamical effects related to baroclinic instabilities, we will show that this model reproduces the main qualitative features of the velocity profiles.

An interesting property of two dimensional turbulent flows is their inverse energy transfer from small scales to large scales, sometimes through a cascade among scales, but much more often through a direct transfer from small scale to large scale mediated by the large scale flow. This inverse
energy transfer is responsible for the self organization of the flow
into large scale coherent structures that may evolve much slower than
the eddies. Among those structures, giant vortices and zonal jets
have raised strong interest in the scientific community. The $\beta$ effect favors the formation of jets, but without $\beta$ effect both jets and vortices can be observed in numerical simulations ~\citep{Sommeria1986,Bouchet_Simonnet_2008,frishman2017jets}. Both structures
are also observed in the atmosphere of gaseous planets \citep{ingersoll1990atmospheric,galperin2014cassini,galperin2001universal}. The computation of statistical equilibrium theory of the two-dimensional Euler and quasi-geostrophic equations  \citep{BouchetVenaille-PhysicsReport}, using large deviation theory,
led to the conclusion that zonal jets as well as large vortices are stable equilibrium states of the flow, and thus natural attractors. However, planetary flows
are continuously damped and forced and a non-equilibrium theory must explain the selection between all such possible attractors.


The exact shape of zonal winds on Jupiter reveals an astonishing asymmetry
between eastward jets and westward jets \citep{porco2003cassini,sanchez2008depth,garci2001study}.
Whereas eastward jets form cusps at their maximum velocity, westward
jets are smoother, close to a parabolic velocity profile. At the same
time, the profile of potential vorticity (PV) looks like ``staircases''
\citep{Dritschel_McIntyre_2008JAtS}, and all those prominent features
are well reproduced in direct numerical simulations of $\beta$ plane turbulence. One could 'postulate'
a potential vorticity staircase profile and derive the corresponding mean flow \citep{Dritschel_McIntyre_2008JAtS}. This exercice is very enlightening, as it roughly relates jet spacing to other flow properties. Nevertheless, the physical mechanism leading to the staircase profile remains unclear. Moreover as our discussion will clearly show, the potential vorticty staircase is just a useful idealised approximation: the actual jet profile will depend on the control parameters, for instance friction, force spectrum, and $\beta$.

Starting from the stochastic barotropic beta plane model, our aim is to derive simple general relations for the jet velocity profile.  A promising nonequilibrium
statistical theory explaining jet formation is the stochastic structural stability (S3T) theory
\citep{Farrel_Ioannou,Farrell_Ioannou_JAS_2007} or the closely related
second order cumulant expansion theory (CE2) \citep{Marston_Conover_Schneider_JAS2008}.
The key ingredient in those theories is to neglect eddy-eddy interactions,
keeping only the interaction between eddies and the mean flow. With this quasilinear approximation, there is no inverse energy cascade
in Fourier space anymore, and the inverse energy flux goes through interactions with the mean flow. This may be relevant only when the inverse energy cascade flux are negligible. The flow governed by the S3T equations, with or without phenomenological added stochastic forcing, produces
spontaneous emergence and equilibration of zonal jets \citep{BakasIoannou2013SSST,constantinou2012emergence}
which velocity profiles reproduce quite well the main features of
jets obtained in rotating-tank experiments \citep{read2004jupiter},
numerical experiments \citep{vallis1993generation,williams1978planetary}
or in the atmosphere of gaseous planets.

The question of why and when this quasilinear approximation should give such good results
has been addressed in \citep{Bouchet_Nardini_Tangarife_2013_Kinetic_JStatPhys}.
The main result is that the quasilinear approximation is self-consistent in the inertial limit of weak stochastic forcing and dissipation, when the inertial time scale is much smaller than the spin up or spin down time scale. For Jupiter the inertial time scale is of order of a day or a month, while the spin up or spin down time scale, related to dissipative phenomena (radiative balance on Jupiter) may be of order of ten years (see e.g \citep{porco2003cassini}). Moreover it follows from \citep{Bouchet_Nardini_Tangarife_2013_Kinetic_JStatPhys} analysis that the quasilinear equations are expected to be valid above a crossover scale, that tends to zero in the limit of weak stochastic forces and dissipation limit. Using this justified approximation in the inertial limit, it is then possible to write a
\emph{closed} equation for the evolution of the mean velocity. 

If we assume that all the energy injected by the force is locally transferred to the mean flow, we obtain
\begin{equation}
\left\langle uv\right\rangle =\frac{\epsilon}{U'},\label{eq:result}
\end{equation}
where $U$ is the mean zonal velocity profile, $U'$ its derivative with respect to the South-North coordinate $y$, $\left\langle uv\right\rangle$ the Reynolds stress, and $\epsilon$ the energy injection rate per unit of mass. Such a formula for the Reynolds stress might give a closed equation for the zonal jet, and is consequently very appealing. This expression is very similar to the one discussed in \citep{laurie2014universal} for a vortex without $\beta$ effect. In this paper this formula was obtained by neglecting the pressure term and the cubic terms in the energy balance relation, without justification. A more general formula, taking into account possible small scale dissipation, was actually obtained previously by \citep{srinivasan2014reynolds}, through explicit computation assuming a constant shear flow $U'=Cst$. A similar result also holds in the case of dipoles for the 2D Navier--Stokes equations \citep{kolokolov2016structure,kolokolov2016velocity}. Through numerical computations \citep{laurie2014universal} have shown that the analogous result for the 2D Navier--Stokes equations  actually predicts correctly the velocity profile in a restricted part of the domain, far from the core of the vortex and far from the flow separatrix. \citep{kolokolov2016structure} give scaling arguments to show in which domain of the flow the theoretical expression for the velocity profile is expected to hold. 

In section \ref{sec:Inertial-small-scale}, following preliminary results in \citep{woillez2017theoretical}, we prove that equation (\ref{eq:result}) can be deduced as a consequence of the two limits of weak forces and dissipation on one hand, and of small scale forcing on the other hand. This first result justifies equation (\ref{eq:result}) and clarifies the required hypothesis. By contrast with our previous work \citep{woillez2017theoretical}, in the present paper we discuss completely the mathematical justification when taking the limit of small scale forces before the inertial limit of weak forces and dissipation. The other order for these limits is way simpler mathematically, but is not relevant for turbulent flows. 

In section \ref{sub:predictions}, we use result (\ref{eq:result}) to write a closed equation for the mean velocity profile $U$. We solve it for the resulting stationary profile. With such an equation, the stationary profile diverges at some finite latitude. We thus conclude that the appealing formula (\ref{eq:result})
is valid only far from the jet tips, where $U'$ does vanish. A more refined analysis is required to deal with the zonal jet velocity
extrema.

In section \ref{sub:Cusps-for-eastward}, using Laplace transform
tools, we derive an equation for the Reynolds stress divergence
in the inertial limit. Taking afterwards the small-scale forcing limit,
we give a set of equations that describes the zonal velocity extremum
of the eastward jet. Although the full numerical calculation of the
solution is avoided, we give some arguments to show that this set
of equations leads to the formation of a ``cusp'' of typical size
$\frac{1}{K}$ where $\frac{1}{K}$  is a typical scale of the stochastic
forcing. We explain that this cusp has no universal shape: it depends on the stochastic forcing spectrum
and on the dissipative mechanism. Yet we derive a relation, valid when viscous phenomena are negligible at the size of the cusp, that
relates the curvature of the cusp to the maximal velocity $U(y_{cr})$.
It writes
\begin{equation}
U(y_{cr})U''(y_{cr})=-\frac{\epsilon K^{2}}{r},\label{eq:result2}
\end{equation}
where $r$ is the linear friction coefficient in $s^{-1}$ (see equation (\ref{eq:curvature})). Remarkably this relation does not depends on the forcing spectrum, but just on $\epsilon$.

On the contrary, the westward jet cannot form this cusp because it
would violate the Rayleigh-Kuo criterion of stability. In section \ref{sub:Computation-of-Reynold's}, we derive a self-consistent equation for the westward jet extrema.
We explain that an instability can develop at the extremum, and how it stops the westward jet growth such
that the zonal flow form a parabolic profile of curvature about $\beta$. This is compatible with the classical idea of barotropic adjustment. We also demonstrate that the appearance of neutral modes (modified Rossby waves) is not sufficient to arrest the growth of the jet extremum velocity, and that a marginal and transient instability is necessary.

\section{Reynolds stresses from energy, enstrophy, and pseudomomentum balances\label{sec:Reynolds-stresses-from}}

\subsection{The stochastic barotropic $\beta$ plane model}

We start from the equations for a barotropic flow on a periodic beta
plane with stochastic forces 
\begin{eqnarray}
\partial_{t}\Omega+V.\nabla\Omega & = & -r\Omega-\beta_{d}V_{y}+\sqrt{2\epsilon}\eta\label{eq:Navier-stokes}\\
\nabla.V & = & 0,\nonumber 
\end{eqnarray}
where the vorticity $\Omega:=\left(\nabla\wedge V\right).e_z$ is the curl of the two dimensional velocity field $V:=\begin{pmatrix}V_{x}\\
V_{y}
\end{pmatrix}$, $r$ models a linear friction, $x$ and $y$ are the East--West and North--South coordinates respectively, $\beta_d$ the Coriolis parameter, and $\eta$ is a stochastic force
that we assume white in time: $\mathbb{E}\left[\eta(x,y,t)\eta(x',y',t')\right]=\delta(t-t')C_{d}(x-x',y-y'')$. We assume that $C_{d}$
is statistically homogeneous such that it depends only on the difference
$x-x'$ and $y-y'$. We choose a particular normalization for the correlation
function $C_{d}$, such that $\epsilon$ is the energy injection rate per unit mass: $\epsilon$ has dimensions $m^{2}s^{-3}$.
In the following, we will always assume that there is no direct energy
injection in the zonal velocity profile, i.e that $\frac{1}{L_{x}}\int{\rm d}x\eta(r,t)=0.$

Nondimensional equations and nondimensional numbers are the clearest way to determine the flow regime.  We choose here
to set temporal and spatial units such that the mean kinetic energy
is 1, and $L_{x}=1$ (please see \citep{Bouchet_Nardini_Tangarife_2013_Kinetic_JStatPhys}
for more details, or (\cite{Bouchet_Nardini_Tangarife_2016_kinetic_Zonal_Jets}
page 2-3) for comparison with other common nondimensionalizations of
the stochastic barotropic equations). For simplicity, we use the same notations for the dimensional and nondimensional velocity and vorticity. The nondimensional equations are 
\begin{eqnarray}
\partial_{t}\Omega+V.\nabla\Omega & = & -\alpha\Omega-\beta V_{y}+\sqrt{2\alpha}\eta,\label{eq:Navier-stokes adim}\\
\nabla. V & = & 0.\nonumber 
\end{eqnarray}
Now $\alpha=L\sqrt{r^{3}/\epsilon}$ is a nondimensional parameter
although we will often refer to it as the ``friction''. $\beta=\sqrt{r/\epsilon}L^{2}\beta_{d}$
is the new nondimensional Coriolis parameter, while $\beta_{d}$ is
the dimensional one. We note that $\beta=L^{2}/L_{R}^{2}$, where
$L_{R}=\left({\epsilon/r\beta_{d}^{2}}\right)^{1/4}$ is the Rhines scale.
The zonostrophy index used in many references would be $R_{\beta}=\beta^{1/10}\epsilon^{1/20}r^{-1/4}$.
We find that $\alpha\propto(R_{\beta})^{-5}$, which implies that
the inertial limit of vanishing $\alpha$ corresponds to the limit of large
$R_{\beta}$. Let $C(r)$ be the nondimensional expression of the
noise correlation function $C_{d}(r)$. We denote $\hat{C}_{k,l}$ the Fourier coefficients
of $C$,
\begin{equation}
C(x,y):=\underset{k,l}{\sum}\hat{C}_{k,l}e^{ikx+ily},\label{eq:Correlation_Fourier}
\end{equation}
and $K^{2}=k^{2}+l^{2}$. As a correlation function,
$C$ is a definite positive function and as a conseuence $\hat{C}_{k,l}$ is real
and positive. Moreover, if we assume the symmetry $x\rightarrow-x$
and $y\rightarrow-y$, the function $\hat{C}_{k,l}$ is symmetric
with respect to  $k\rightarrow-k$ and $l\rightarrow-l$. The constrain that the
mean kinetic energy is one writes
\begin{equation}
\frac{1}{2}\iint{\rm d}k{\rm d}l\frac{\hat{C}_{k,l}}{K^{2}}=1.
\end{equation} 
From now on, the computations will be done with nondimensional quantities.
If we want to write a result in its dimensional formulation, we will
reintroduce $\left[\frac{\epsilon}{r}\right]=m^{2}.s^{-2}$ and $[L_{x}]=m$.

We separate the flow $V$ in two parts, $V(r,t)=U(y,t)e_{x}+\begin{pmatrix}u(r,t)\\
v(r,t)
\end{pmatrix}$. The mean velocity $Ue_{x}=\left\langle V\right\rangle $ is defined
as the zonal and stochastic average of the velocity field. More precisely,
we assume that the mean flow is parallel and we take $\left\langle V\right\rangle=\frac{1}{L_{x}}\int{\rm d}x\mathbb{E}[V(x,y)]$.
In the following, the bracket $\left\langle \right\rangle $ will
be used for both the zonal and stochastic averages. The vorticity then
separates in $\Omega(r,t)=-U'(y,t)+\omega(r,t)$, where the prime
denotes the derivative with respect to $y$. We will refer to $U$ indifferently as the \emph{mean
flow} or \emph{zonal flow}. 

Using this decomposition and the continuity equation, we can obtain
an equation for the zonal component of the vorticity, and then integrate
over $y$ to get the equation for the mean velocity $U$ 
\begin{equation}
\partial_{t}U+\partial_{y}\left\langle uv\right\rangle =-rU.\label{eq:meanvelocity}
\end{equation}
Equation (\ref{eq:meanvelocity}) shows that the mean flow
is forced by the divergence of the Reynolds stress $\partial_{y}\left\langle uv\right\rangle $.
In order to reach an equilibrium, this latter term has to balance
the dissipation coming from linear friction.

\subsection{Quasilinear approximation and pseudomomentum balance}

In this section we define the quasilinear approximation. We recall basic concepts about linear dynamics and the relation between stochastic and deterministic linear dynamics.

In the following we are interested in the small $\alpha$ regime (the inertial or weak forces and dissipation regime).
In the limit where $\alpha$ goes to zero in equation (\ref{eq:Navier-stokes adim}), we can neglect the nonlinear eddy-eddy interactions. This approximation is called the \emph{quasilinear approximation}. The quasilinear approximation has been shown self-consistent by \citep{Bouchet_Nardini_Tangarife_2013_Kinetic_JStatPhys} with some assumptions on the profile $U$ (stability, no zero modes). We will not develop the full justification of the quasilinear approximation
here, the interested reader is referred to \citep{Bouchet_Nardini_Tangarife_2013_Kinetic_JStatPhys}.
Let us simply recall the heuristic ideas leading to the quasilinear
approximation. First, we notice that the strength of the noise is
of order $\sqrt{\alpha}$. As fluctuations are sheared and transferred
to the largest scales on a timescale of order one, this is a natural
hypothesis to expect fluctuations $(u,v)$ to be of the same order.
This was proven to be self-consistent in \citep{Bouchet_Nardini_Tangarife_2013_Kinetic_JStatPhys}.
We make the rescaling $(u,v):=\sqrt{2\alpha}(u',v')$ in equation
(\ref{eq:Navier-stokes adim}), and we omit the prime in the following for clarity. The eddy-eddy interaction terms are of
order $\alpha^{\frac{3}{2}}$, and can then be neglected. We are left
with the set of equations 
\begin{equation}
\partial_{t}U=-\alpha\left[\partial_{y}\left\langle uv\right\rangle +U\right]\label{eq:quasilinear}
\end{equation}
\begin{equation}
\partial_{t}\omega+U\partial_{x}\omega+(\beta-U'')v=-\alpha\omega+\eta\label{eq:quasilinear2}
\end{equation}
where  $\omega=\partial_{x}v-\partial_{y}u=\triangle\psi$ is the rescaled 
vorticity fluctuations. Equation (\ref{eq:quasilinear}) shows
that the typical timescale for the evolution of the mean flow $U$
is $\frac{1}{\alpha}$. By contrast, equation (\ref{eq:quasilinear2}) shows that the timescale for eddy dynamics is of order one. Using
this timescale separation, we will consider that $U$ is a constant
field in the second equation (\ref{eq:quasilinear2}), and we will
solve $\omega(t)$ for a given profile $U$. Once $U$ is considered as given, the eddy equation
is linear. This time scale separation is observed for example on Jupiter where the typical
time of eddies evolution ranges from few days to few weeks whereas
significant changes in the mean flow are only detected over decades
(see e.g \citep{porco2003cassini}).\\

Without forces and dissipation, the quasilinear equations conserve energy and enstrophy as do the
full barotropic flow equations. One of the key relations
we will use in this paper comes from the fluctuation enstrophy balance
\[
\frac{1}{2}\partial_{t}\left\langle \omega^{2}\right\rangle -(\beta-U'')\partial_{y}\left\langle uv\right\rangle =-\alpha\left\langle \omega^{2}\right\rangle +\frac{1}{2}C(0).
\]
where we have used 
\begin{equation}
\left\langle v\omega\right\rangle =-\partial_{y}\left\langle uv\right\rangle,\label{eq:tenseur2}
\end{equation}
which a consequence of incompressibility.

As a consequence, if $U''-\beta$ has a constant sign in the flow, without forces and dissipation, the left-hand side of equation
(\ref{eq:quasilinear2}) conserves the pseudomomentum $\int\frac{\left\langle \omega^{2}\right\rangle }{U''-\beta}\rm{d}y$. The pseudomomentum does not allow any instability to occur and the
flow is stable. This is called the Rayleigh-Kuo criterion for stability
of shear flows. If $U''-\beta$ vanishes somewhere in the flow, an
instability may or may not exist. The fact that $U''-\beta$ vanishes
is a necessary condition for instability, not a sufficient one.  

As we assume a timescale separation between the zonal flow and fluctuation
dynamics, we are interested in the long-term behavior of the Reynolds stress $\partial_{y}\left\langle uv\right\rangle$. When the vorticity fluctuations $\omega$ reach its stationary
distribution, we have the relation
\begin{equation}
\partial_{y}\left\langle uv\right\rangle = -\frac{1}{U''-\beta}\left[\alpha\left\langle \omega^{2}\right\rangle -\frac{1}{2}C(0)\right].\label{eq:tenseur}
\end{equation}
This equation for the Reynolds stress will be extremely useful.

We now take the Fourier transform of (\ref{eq:quasilinear2}) in $x$: $\omega_{k}(y):=\frac{1}{L_{x}}\int{\rm d}x\omega(x,y)e^{-ikx}$
with $k$ taking the values $\frac{2\pi}{L_{x}}n$, $n$ is an integer. We also use the linearity
to express the solution as the sum of particular solutions for independent stochastic forcings $\eta_{l}(y,t)$. We denote $\omega_{k,l}(y,t)$ the solution of 
\[
\partial_{t}\omega_{k,l}+L_{k}[\omega_{k,l}]=-\alpha\omega_{k,l}+\eta_{l},\label{eq:quasilinearcomplex}
\]
where 
\begin{equation}
L_{k}[\omega_{k,l}]=ikU\omega_{k,l}+ik(\beta-U'')\psi_{k,l},\label{eq:Lk}
\end{equation}
and where $\eta_{l}$ is a Gaussian white noise with correlations \textbf{\emph{$\mathbb{E}\left[\eta_{l}(y,t)\eta_{l}(y',t)\right]=e^{il(y-y')}\delta(t-t')$}}. 
From equation (\ref{eq:tenseur}) we obtain
\begin{equation}
\partial_{y}\left\langle uv\right\rangle = -\frac{1}{U''-\beta}\underset{k,l}{\sum}\frac{\hat{C}_{k,l}}{2}\left[2\alpha\left\langle |\omega_{k,l}|^{2}\right\rangle -1\right],\label{eq:tenseur decompose}
\end{equation}
where the positive constants $\hat{C}_{k,l}$ are defined by (\ref{eq:Correlation_Fourier}). We stress that\textbf{ }in this formula the bracket $\left\langle |\omega_{k,l}|^{2}\right\rangle $
denotes a stochastic average, because the zonal average is already taken into account by the sum over all vectors $k$. 

We now proceed to a further simplification, showing that the stochastic eddy dynamics can be computed from a set of deterministic linear problems, following\emph{ }\citep{Bouchet_Nardini_Tangarife_2013_Kinetic_JStatPhys}. We use that equation (\ref{eq:Lk}) is a linear operator for a given $U$ and that the
noise $\eta_{k,l}$ is white in time and has an exponential correlation
function \textbf{$c_{l}(y)=e^{ily}$} to express the stationary average
$\left\langle |\omega_{k,l}|^{2}\right\rangle $ as 
\begin{equation}
\left\langle |\omega_{k,l}|^{2}\right\rangle =\int_{-\infty}^{0}{\rm d}t~e^{2\alpha t}\left|e^{tL_{k}}[c_{l}]\right|^{2},\label{eq:omegacarre}
\end{equation}
where $e^{tL_{k}}[c_{l}]$ is the solution at time $t$ of the \emph{deterministic equation }
\[
\partial_{t}\omega_{d}+L_{k}[\omega_{d}]=0, \label{eq:deterministic}
\]
 with initial condition $c_{l}:=y\rightarrow e^{ily}$. 
 
Equations (\ref{eq:tenseur decompose}), (\ref{eq:omegacarre}) and (\ref{eq:quasilinear}) give a way to compute Reynolds stresses and the velocity profile $U$ from classical and much studied deterministic hydrodynamic problems. Equation (\ref{eq:deterministic}) is however tricky and has no simple explicit expression in the general case. We will be able to get explicit result only in asymptotic regimes.

\subsection{Simplifications in asymptotic regimes}

Expression (\ref{eq:omegacarre}) is still complicated because to
get explicit results, it requires\textbf{ }to know the behavior of
the solution \textbf{$e^{tL_{k}}[c_{l}]$} up to times of order $\frac{1}{\alpha}$.
Two parameters can be used to further simplify the problem, the vector
$\mathbf{k}=(k,l)$ and the damping $\alpha$. We denote $K:=|\mathbf{k}|$.
We will be interested both in the regime $K\rightarrow\infty$
and $\alpha\rightarrow0$. The large $K$ regime is a small scale forcing regime.

The inertial limit $\alpha\rightarrow0$ is the most difficult one, because
turbulence can develop on a very long time. But the inertial limit is also the most
interesting from a physical point of view because it corresponds to
fully turbulent regimes, and is the most relevant one to describe Jupiter's atmosphere. In this section we prove that equation (\ref{eq:tenseur decompose}) and (\ref{eq:omegacarre}) can be further simplified and computed through the asymptotic behavior of a linear dynamics without dissipation, in the inertial regime.

The result in
the inertial regime crucially depends on whether
the deterministic equation without dissipation 
\begin{equation}
\partial_{t}\omega_{d}+ikU\omega_{d}+ik(\beta-U'')\psi_{d}=0\label{eq:deterministic}
\end{equation}
with initial condition $\omega_d(y,0)=c_l$, sustains neutral modes or not. A neutral mode is defined as a solution
of this equation of the form $\omega_d(y,t)=\xi^{a}(y)e^{ic_{a}t}$
where $c_{a}$ is a real constant. It is also sometimes called ``modified
Rossby waves'' in this context, when the jet velocity is nonzero. Two cases can be encountered.
In the first case, without neutral modes, \citep{Bouchet_Morita_2010PhyD}
have shown that $\omega_{d}$
behaves asymptotically for long time as $\omega_{d}(y,t)\underset{t\rightarrow\infty}{\sim}\omega^{\infty}(y)e^{ikUt}$,
even for non monotonous velocity profiles $U$. Please note that we should write $\omega_{kl}^{\infty}(y)$ because the asymptotic limit of $\omega_d$ depends on the wavevector, but we choose to omit the indices $k,l$ for clarity. Moreover \citep{Bouchet_Morita_2010PhyD} give
a method to compute this $\omega^{\infty}$ using Laplace transform
tools. In this case, in the limit $\alpha\rightarrow0$, the Reynolds stress writes 
\begin{equation}
\partial_{y}\left\langle uv\right\rangle  = -\frac{1}{U''-\beta}\underset{k,l}{\sum}\frac{\hat{C}_{k,l}}{2}\left[|\omega^{\infty}|^{2}-1\right].\label{eq:inertial}
\end{equation}
We give the full justification of this result in appendix \ref{sec:The-Reynold's-stress}.

In the second case, with neutral modes, we have to modify expression (\ref{eq:inertial})
to take into account the presence of modes. Again, we leave the technical
details to appendice \ref{sec:The-Reynold's-stress} and we give the
final result 
\begin{equation}
\partial_{y}\left\langle uv\right\rangle = -\frac{1}{U''-\beta}\underset{k,l}{\sum}\frac{\hat{C}_{k,l}}{2}\left[|\tilde{\omega}^{\infty}|^{2}-1+\underset{modes~a}{\sum}|\omega^{a}|^{2}\right].\label{eq:inertialmodes}
\end{equation}
This result means that we have to project first the initial condition
$c_{l}$ over the modes labeled by $a$. The component over the $a$
mode gives the term $\omega^{a}(y)$. This new terms are related
to the wave pseudomomentum balance. Then we compute the asymptotic
solution $\tilde{\omega}^{\infty}$ of (\ref{eq:deterministic})
using as initial condition not $c_{l}$ but $c_{l}-\sum\omega^{a}$.
Briefly speaking, a first reason why there are no cross terms of the
form $\omega^{a}\tilde{\omega}$ between modes and the remaining part
of the spectrum is because the frequencies $c_{a}$ of the modes are
always outside of the range of $U$ as shown by \citep{drazin1982rossby,pedlosky1964stability}. The cross terms have an oscillatory
part of frequency $\frac{1}{\alpha}(c_{a}-U)$ that gives a vanishing
contribution in the small $\alpha$ limit.

Both formulas (\ref{eq:inertial}-\ref{eq:inertialmodes}) are independent from $\alpha$  in the limit of vanishing $\alpha$. This is a non-trivial result. The formulas will be
used to study the inertial limit in section \ref{sec:Inertial-small-scale}.

\section{Explicit velocity profile in the inertial and small scale forcing regime\label{sec:Inertial-small-scale}}

Observations collected by the Gallileo and Cassini probes (see \citep{porco2003cassini}
and others) allows to estimate typical values for $K$ and $\alpha$.
$1/K$ is the forcing length scale. It can be estimated to be or order $1000$ km, the typical size of cyclones due to convective activity in Jupiter's
troposphere. The dissipation on Jupiter involves different mechanisms, which are roughly modelled by our linear friction. What could is a relevant typical
timescale for dissipation is not obvious. Based on Jupiter observations, many authors \citep{porco2003cassini,vasavada2005jovian,salyk2006interaction} consider a large scale dissipation time $1/r$ of the order of a few years. To compute an order of magnitude for the non dimensional parameter $\alpha$, we have chosen $1/r=5$ years. $U$  is easily estimated from the observations, and the Coriolis parameter
$\beta$ is easily computed from the rotation rate of Jupiter. After non-dimensionalisation, following the discussion in the previous section, we estimate the orders of magnitude for $K\sim10$ and $\alpha\sim10^{-3}$. Jupiter
is thus in the asymptotic regime $\alpha\rightarrow0$
and $K\rightarrow+\infty$. Both limits do not necessarily commute,
we thus have to be careful which limit we are going to take first. The turbulent nature of the dynamics at the forcing scale suggests the limit $\alpha\rightarrow0$ first, and then $K\rightarrow+\infty$.

\subsection{Computation of the long-time limit of eddy vorticity\label{sub:Computation-of}}

In the following and until section \ref{sub:Computation-of-Reynold's},
we assume there are no Rossby waves in the flow. It has been shown
long ago that those waves travel in a barotropic flow at a velocity $c<U_{min}$ \citep{drazin1982rossby,pedlosky1964stability}.
In section \ref{sub:Computation-of-Reynold's},
we will explain how we can compute Rossby waves in case of a parabolic profile.

In this subsection, we summarize the main result obtained by Bouchet
and Morita \citep{Bouchet_Morita_2010PhyD} that allows us to compute
the function $\omega^{\infty}$  appearing in (\ref{eq:inertial}). $\omega^{\infty}$ gives then an easy access to the small $\alpha$
limit, independently of the large $K$ limit.

We start from equation (\ref{eq:deterministic}) that describes the
linear evolution of a perturbation $\omega(y,t)\mbox{e}^{ikx}$ of
meridional wave number $k$, and with streamfunction $\psi(y,t){\rm e}^{ikx}$. In the following, we will stop using the subscript $d$ for the deterministic solution $\omega_{d}$ in (\ref{eq:deterministic}).
We introduce the function $\varphi_{\epsilon}(y,c)$ which is the
Laplace transform of the stream function $\psi(y,t)$ i.e $\varphi_{\epsilon}(c):=\int_{0}^{\infty}{\rm d}t\psi(y,t)e^{ik(c+i\epsilon)t}$. To avoid any confusion, we stress that in this section, $\epsilon$
will always denote a small parameter and not the energy
injection rate. The Laplace transform $\varphi_{\epsilon}$ is well
defined for any non zero value of the real variable $\epsilon$ with
a strictly positive product $k\epsilon$. $c$ has to be understood
as the phase speed of the wave, and $k\epsilon$ is the
exponential growth rate of the wave. Note that $k\epsilon$ exactly corresponds to a linear friction in equation (\ref{eq:deterministic}), such that the inertial limit $\alpha\rightarrow0$ is equivalent to the limit $\epsilon\rightarrow0$.  The equation for $\varphi_{\epsilon}$
is 
\begin{equation}
\left(\frac{d^{2}}{dy^{2}}-k^{2}\right)\varphi_{\epsilon}(y,c)+\frac{\beta-U''(y)}{U(y)-c-i\epsilon}\varphi_{\epsilon}(y,c)=\frac{\omega(y,0)}{ik(U(y)-c-i\epsilon)},\label{eq:inhomogeneous rayleigh}
\end{equation}
(see \citep{Bouchet_Morita_2010PhyD}), with vanishing boundary conditions at
infinity. We do not have an infinite flow in the $y$ direction, but
the properties of the flow become local for large $K$. The choice
to take vanishing boundary conditions at infinity is done for convenience
and it is expected that this particular choice does not modify the
physical behavior of the perturbation.

For all $\epsilon>0$ the function $\varphi_{\epsilon}$ is well defined.
The inhomogeneous Rayleigh equation (\ref{eq:inhomogeneous rayleigh})
is singular for $\epsilon=0$ at any critical point (or critical
layer) $y_{c}$ such that the zonal velocity is equal to the phase
speed: $U(y_{c})=c$. One can show that $\varphi_{\epsilon}$
has a limit denoted $\varphi_{+}$ when $\epsilon$ goes to zero.
The function $\omega^{\infty}$ is then given by 
\begin{equation}
\omega^{\infty}(y)=ik(U''(y)-\beta)\varphi_{+}(y,U(y))+\omega(y,0),\label{eq:omegainfini}
\end{equation}
see \citep{Bouchet_Morita_2010PhyD}. The function $\omega^{\infty}$
depends on the Laplace transform of the stream function but for a
phase velocity $c$ equal to the zonal velocity at latitude\textbf{
$y$}. From a mathematical point of view, it corresponds to the value
of $\varphi_{+}$ exactly at its singularity. The singularity in equation
(\ref{eq:inhomogeneous rayleigh}) is of degree one (proportional
to $\frac{1}{y}$) except at the extrema of the jets where it is of
degree two. A singularity of order two would create a divergence for
the solution, but it happens that the numerator in (\ref{eq:inhomogeneous rayleigh})
vanishes at such points and the solution is still defined at the extrema
of a jet. A nontrivial consequence of that is 
\[
\omega^{\infty}(y_{cr})=0
\]
at all critical latitudes $y_{cr}$ where $U'(y_{cr})=0.$ This result,
called depletion of vorticity fluctuation at the jet critical points
in \citep{Bouchet_Morita_2010PhyD}, has important physical consequences
that influence the dynamics of a jet.

As described in \citep{Bouchet_Morita_2010PhyD}, using formula (\ref{eq:inhomogeneous rayleigh})
and (\ref{eq:omegainfini}), one can numerically compute the function
$\omega^{\infty}$ : we first have to solve a set of boundary value
problems for ordinary differential equations parameterized by $c$
and $\epsilon$ to obtain a solution family\textbf{ $\varphi_{\epsilon}(c)$}.
Then we evaluate, for small enough $\epsilon$ each solution $\varphi_{\epsilon}(c)$
at the value $y_{c}$ satisfying $U(y_{c})=c$. This method is much
faster and has less numerical cost than computing the long time evolution
of the partial differential equation (\ref{eq:deterministic}). We
use this method in the following of this section.

\subsection{Limit of small scale forcing for monotonic profiles, and explicit
expression of the Reynolds stress}\label{sub:limit of small scale}

Again, we assume there are no Rossby waves. We will now consider the
limit of small scale forcing $K\rightarrow\infty$. The calculations
are rather technical and can be skipped in the first lecture. The result of this section is equation (\ref{eq:formule1 adim})

We start from equation (\ref{eq:inhomogeneous rayleigh}) that describes
the inertial behavior of a deterministic evolution of a perturbation
$\omega(y,0)$ when $\epsilon$ vanishes. Using the Green function
$H_{k}(y)$ of $(\partial_{y}^{2}-k^{2})$ we write 
\[
\varphi_{\epsilon}(y,c)=\left(U''(y)-\beta\right)\int{\rm d}y'H_{k}(y')\frac{\varphi_{\epsilon}(y-y',c)}{U(y-y')-c-i\epsilon}+\int{\rm d}y'H_{k}(y')\frac{\omega(y-y',0)}{ik(U(y-y')-c-i\epsilon)}.
\]
Now we make the change of variable $Y=ky'$ . The Green function has
the scaling $H_{k}(y'):=-\frac{1}{2k}H_{0}(Y)$. Recalling that $\varphi_{+}(y,c)=\lim_{\epsilon\downarrow0}\varphi_{\epsilon}(y,c)$,
it follows 
\begin{multline}
\varphi_{+}(y,c)=-\frac{\left(U''(y)-\beta\right)}{2k^{2}}\underset{\epsilon\rightarrow0}{\lim}\int{\rm d}YH_{0}(Y)\frac{\varphi_{\epsilon}(y-\frac{Y}{k},c)}{U(y-\frac{Y}{k})-c-i\epsilon}\\
-\frac{1}{2ik^{3}}\underset{\epsilon\rightarrow0}{\lim}\int{\rm d}YH_{0}(Y)\frac{\omega(y-\frac{Y}{k},0)}{U(y-\frac{Y}{k})-c-i\epsilon}.\label{eq:phiplus}
\end{multline}
Please note that we are making the asumption that $\frac{l}{k}:=\tan\theta$
is finite and thus $K\rightarrow\infty$ implies $k\rightarrow\infty$.
Let us remind here that it is crucial to take the limit $\epsilon\rightarrow0$
first before $K\rightarrow\infty$ because $\epsilon$ exactly plays
the role of the nondimensional linear friction $\alpha$. If we want
to study the inertial regime, we have to take first a vanishing friction
limit.

Consider now the magnitude of both terms in the right-hand side of
(\ref{eq:phiplus}). We have a term depending on $\varphi_{\epsilon}$ and
another depending on the initial condition $\omega(y,0)$. The initial
condition is of order 1, and then the second term will be of order
$\frac{1}{k^{3}}$. As a consequence, the first term in the asymptotic
expansion of $\varphi_{+}$ will be of order $\frac{1}{k^{3}}$. The
first term in the right-hand side of (\ref{eq:phiplus}) gives the
next order of the asymptotic expansion and is thus negligible. We
write 
\begin{equation}
\varphi_{+}(y,c)\underset{K\rightarrow\infty}{\sim}-\frac{1}{2ik^{3}}\underset{\epsilon\rightarrow0}{\lim}\int{\rm d}YH_{0}(Y)\frac{\omega(y-\frac{Y}{k},0)}{U(y-\frac{Y}{k})-c-i\epsilon}.\label{eq:phiplus-asymptotic}
\end{equation}
Combining equations (\ref{eq:omegainfini}) and (\ref{eq:phiplus-asymptotic})
we find that 
\[
|\omega^{\infty}(y)|^{2}\underset{K\rightarrow\infty}{\sim}|\omega(y,0)|^{2}-\frac{U''-\beta}{k^{2}}\mathcal{R}e\left\{ \underset{\epsilon\rightarrow0}{\lim}\int{\rm d}YH_{0}(Y)\frac{\omega^{*}(y,0)\omega(y-\frac{Y}{k},0)}{U(y-\frac{Y}{k})-U(y)-i\epsilon}\right\} .
\]
The final step is to use $\omega(y,0)=e^{ily}$, and $H_{0}(Y)=e^{-|Y|}$.
We use also the Sokhotski\textendash Plemelj formula: $\underset{\epsilon\rightarrow0}{\lim}\frac{1}{x-i\epsilon}=i\pi\delta(x)+\mathcal{P}\left(\frac{1}{x}\right),$
to obtain 
\begin{eqnarray*}
|\omega^{\infty}(y)|^{2} & \underset{K\rightarrow\infty}{\sim} & |\omega(y,0)|^{2}-\frac{U''-\beta}{k^{2}}\mathcal{R}e\left\{ \underset{\epsilon\rightarrow0}{\lim}\int{\rm d}Ye^{-|Y|}\frac{e^{-iY\tan\theta}}{U(y-\frac{Y}{k})-U(y)-i\epsilon}\right\} \\
 & \underset{K\rightarrow\infty}{\sim} & |\omega(y,0)|^{2}-\frac{U''-\beta}{k^{2}}\mathcal{R}e\left\{ i\pi\int{\rm d}Ye^{-|Y|}e^{-iY\tan\theta}\delta\left(U\left(y-\frac{Y}{k}\right)-U(y)\right)\right\} \\
 &  & -\frac{U''-\beta}{k^{2}}\mathcal{R}e\left\{ \mathcal{P}\left\{ \int{\rm d}Ye^{-|Y|}\frac{e^{-iY\tan\theta}}{U(y-\frac{Y}{k})-U(y)}\right\} \right\} \\
 & \underset{K\rightarrow\infty}{\sim} & |\omega(y,0)|^{2}-\frac{U''-\beta}{k^{2}}\mathcal{P}\left\{ \int{\rm d}Ye^{-|Y|}\frac{cos(Y\tan\theta)}{U(y-\frac{Y}{k})-U(y)}\right\} ,
\end{eqnarray*}
where we have used that the term $i\pi\int{\rm d}Ye^{-|Y|}e^{-iY\tan\theta}\delta\left(U\left(y-\frac{Y}{k}\right)-U(y)\right)$
is purely imaginary. Injecting this result in (\ref{eq:inertial})
gives the contribution of one Fourier mode $k,l$ with $\frac{k}{l}=\tan\theta$
to the Reynolds stress divergence 
\[
\mathcal{R}e\left\langle v_{\theta}^{*}\omega_{\theta}\right\rangle \underset{K\rightarrow\infty}{\sim}-\frac{\hat{C}_{k,l}}{2k^{2}}\mathcal{P}\left\{ \int{\rm d}Ye^{-|Y|}\frac{\cos(Y\tan\theta)}{U(y-\frac{Y}{k})-U(y)}\right\} .
\]
Some lengthy but straightforward calculations are then required to
show that this expression coincides with $\frac{U''}{U'^{2}}$ (for
an energy injection rate set to one). The computation is discussed in appendix \ref{sec:Equivalence-of-the_limits}.
But we have to do an additional assumption: the asymptotic expansion
is valid only if $\frac{kU'}{U''}\rightarrow\infty$. There should
exist a small region in the vicinity of the extremum $U'=0$ where
the calculation breaks down. The formula can be valid only for strictly
monotonic profiles or for the monotonic part between two extrema of
a jet.\\ 

We have derived the first main result of the present paper,
\begin{equation}
\partial_{y}\left\langle uv\right\rangle =-\frac{U''}{U'^{2}},\label{eq:formule1 adim}
\end{equation}
in the limits $\alpha\rightarrow0$ and $K\rightarrow+\infty$ taken
in this order. With the dimensional physical fields, the result (\ref{eq:formule1 adim})
writes
\begin{equation}
\partial_{y}\left\langle uv\right\rangle =-\epsilon\frac{U''}{U'^{2}},\label{eq:formule1 dim}
\end{equation}
where $\epsilon$ is the energy injection rate.

We have proven that in the limit of vanishing friction and small scale
forcing, we are able to give an explicit expression for the Reynolds
stress divergence that does not depend on the shape of the stochastic
forcing. The result (\ref{eq:formule1 dim}) has been obtained taking
the limit $\alpha\rightarrow0$ first. It has been shown by \cite{woillez2017theoretical}
that the result (\ref{eq:formule1 dim}) can also be recovered taking
the limit $K\rightarrow+\infty$ before $\alpha\rightarrow0$, which
means that both limits do commute in the present case. It is worth
emphasizing that our results are asymptotic results. The behavior
may be really different for finite friction and finite $K$. The work
done in \citep{srinivasan2014reynolds} shows that the shape of the
stochastic forcing matters in the general case.

\subsection{Prediction of the stationary velocity profile}\label{sub:predictions}

With the asymptotic result (\ref{eq:formule1 dim}), we are now able
to derive a close equation for the mean velocity profile $U$. Relation~(\ref{eq:formule1 dim}) together with equation~(\ref{eq:meanvelocity})
gives
\begin{equation}
\partial_{t}U-\frac{\epsilon U"}{U'^{2}}=-rU.\label{eq:close equation velocity}
\end{equation}
From the latter result, we deduce that the stationary velocity profile
$U_{0}$ satisfies the equation
\begin{equation}
\frac{\epsilon U_{0}"}{U_{0}'^{2}}=rU_{0}.\label{eq:stationary1}
\end{equation}
Equation (\ref{eq:stationary1}) surprisingly has a Newtonian structure, it has a first integral that can be interpreted as the sum of a kinetic energy and a potential energy. 
Multiplying both sides by $U_{0}'$ and integrating over $y$ leads to
\begin{equation}
\frac{1}{2}U_{0}^{2}-\frac{\epsilon}{r}\ln\left(\left|U_{0}'\right|\right)=C,\label{eq:particle in potential}
\end{equation}
where $C$ is an integration constant. 

\begin{figure}
\begin{centering}
\includegraphics[height=7cm]{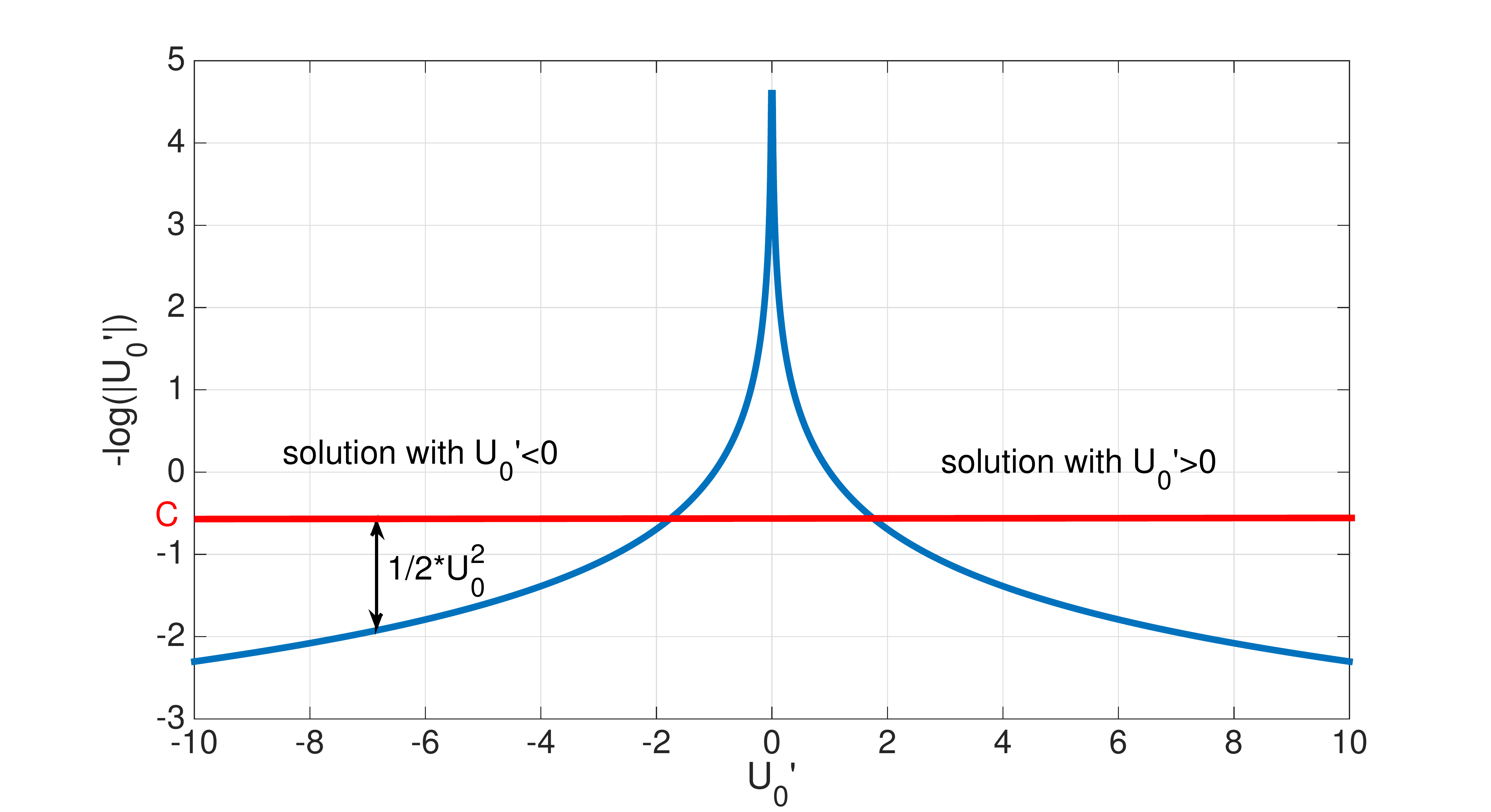}
\end{centering}
\caption{Newtonian structure of the mean flow dynamics.
The stationary zonal flow satisfies a Newtonian equation similarly
to a particle in a potential. Whatever the value of the integration
constant $C$, two classes of solutions exist: one profile has an
increasing velocity, the other one has a decreasing velocity. \label{fig:potential}}
\end{figure}

In equation (\ref{eq:particle in potential}), the function $V(x):=-\frac{\epsilon}{r}\ln\left(\left|x\right|\right)$
plays the role of a potential. The dynamics defined by (\ref{eq:particle in potential})
is completely similar to a particle moving in a potential $V$ with equation
\[
\frac{1}{2}\dot{x}^{2}+V(x)=C.
\]
The only difference is that the roles of
$U$ and $U'$ are exchanged compared to the role of $x$ and $\dot{x}$
for a particle in a potential. The situation is represented in figure~(\ref{fig:potential}).

\begin{figure}
\begin{centering}
\includegraphics[height=9cm]{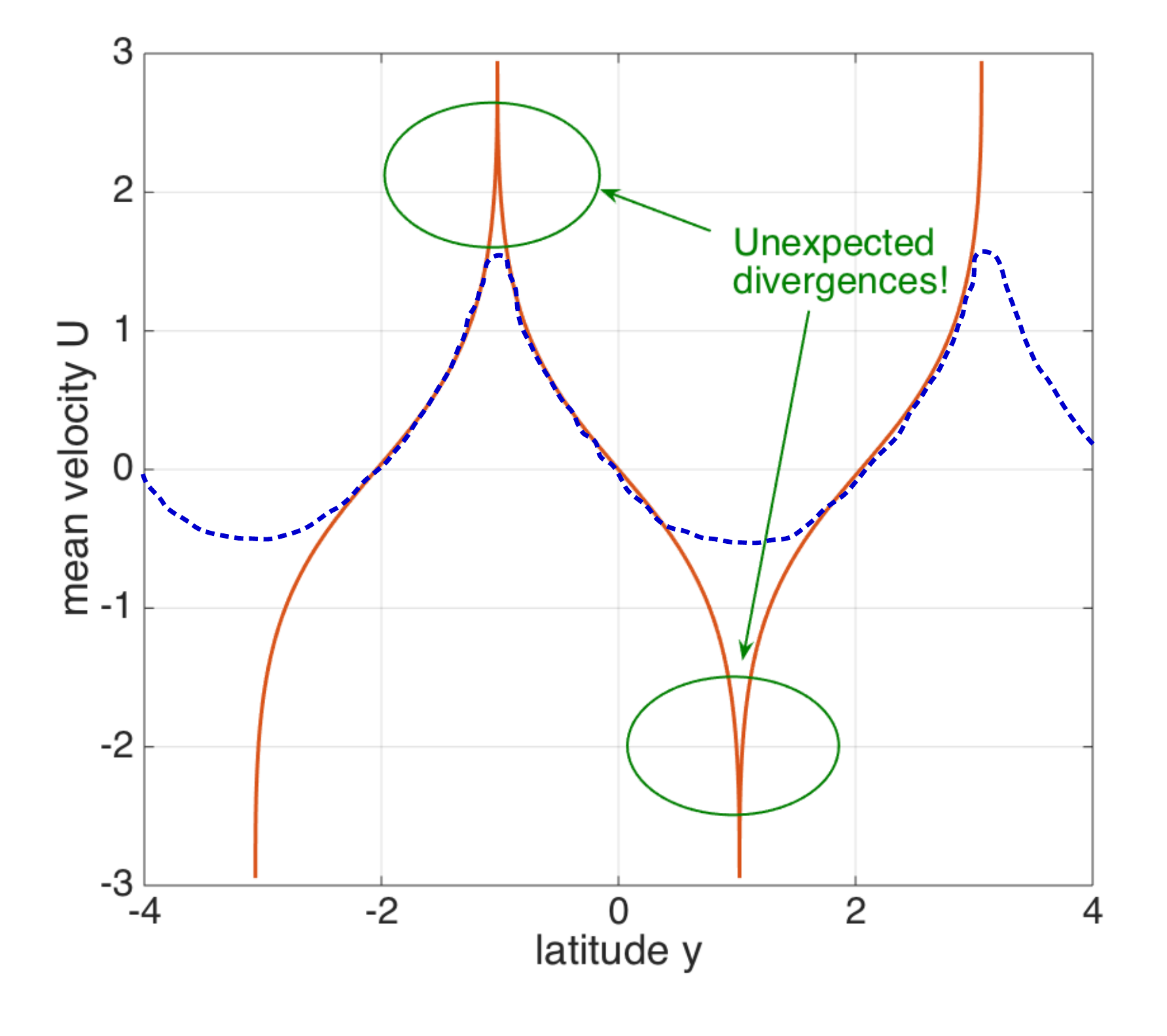}
\par\end{centering}
\caption{Mean velocity profile in the inertial small scale forcing limit.
(Color online) The red curve displays the stationary velocity profile of the zonal flow predicted by equation
(\ref{eq:particle in potential}), in the inertial small scale forcing
limit. Equation (\ref{eq:particle in potential}) predicts symmetric
eastward and westward jets, with diverging values of the velocity
at the extrema. The blue curve displays the qualitative shape of a real profile as observed on Jupiter or in numerical simulations (see e.g. figures~(\ref{fig:The-eastward-jet}) and (\ref{fig: jupiter jets})). Two different regularization mechanisms prevent the jet divergence in the eastward part and in the westward part respectively. \label{fig:diverging profiles}}
\end{figure}

Whatever the value of the constant $C$, the velocity profile $U_{0}$
always diverges. The derivative $U_{0}'$ cannot change sign. There
are two classes of solutions, either solutions with $U_{0}'>0$ or solutions
with $U_{0}'<0$. The two classes of solutions correspond to the two
sides of a jet. The solution of equation (\ref{eq:particle in potential})
is represented in figure~(\ref{fig:diverging profiles}). Equation (\ref{eq:particle in potential})
predicts that zonal jets are composed by a succession of diverging
velocity profiles, with successively increasing and decreasing values
of the velocity. The side of increasing velocity of a jet is totally
independent of the side with decreasing velocity. The velocity profiles
of westward and eastward jets are symmetric, with in both cases a
diverging value of the velocity at the extremum. Such a velocity profile is
of course not realistic because the velocity of zonal winds have
finite values. By contrast, the qualitative shape of a real profile is displayed by the blue curve in figure~(\ref{fig:diverging profiles}). Equation (\ref{eq:particle in potential}) predicts the velocity profile in the intermediate regions of monotonic velocity, away from the jet edges.

The fact that equation (\ref{eq:particle in potential}) predicts
divergent velocity profiles means that some of the hypotheses leading
to the result (\ref{eq:formule1 dim}) are broken at the extrema of
zonal jets. The asymptotic expansion has been obtained using two major
asumptions: first, the limit $\frac{kU'}{U''}\rightarrow\infty$ should
be satisfied, and second the mean flow should be hydrodynamically
stable (no unstable modes in equation (\ref{eq:inhomogeneous rayleigh})).
The first assumption is broken at the eastward extrema of jets, and
the second is broken at the westward extrema. Section \ref{sub:Cusps-for-eastward}
explains the regularization mechanism that creates a cusp at the eastward
extremum, and section \ref{sub:The-question-of} shows that an hydrodynamic
instability stops the growth of the westward jet at the maximal curvature
$\beta$. By taking those physical mechanisms into account, it is
possible to get realistic jets that correspond to the observations
on Jupiter's troposphere. The width of jets is not constrained by equation (\ref{eq:stationary1}).  The typical width of a jet is set by the limit curvature $\beta$ at the westward extremum that imposes a minimal spacing between two consecutive jets.\\

\subsection{Interpretation of eq.(\ref{eq:formule1 dim}) from the energy balance}\label{sub:interpretation}

We now give a very enlightening
interpretation of the result (\ref{eq:formule1 dim}) in terms of
the energy balance. Multiplying both sides of equation (\ref{eq:meanvelocity}), we obtain the energy balance equation for the large scales of the flow
\begin{equation}
\partial_{t}\left(\frac{1}{2}U^{2}\right)+\partial_{y}\left(U\left\langle uv\right\rangle \right)=U'\left\langle uv\right\rangle -rU^{2}.
\label{eq:energy bilan}
\end{equation}
We interpret the different terms in equation (\ref{eq:energy bilan}). $\frac{1}{2}U^{2}$ is the kinetic energy density. The term $\partial_{y}\left(U\left\langle uv\right\rangle \right)$ is a divergence, and thus the quantity $U\left\langle uv\right\rangle$ can be interpreted as the spatial energy flux at large scales. Energy is dissipated by the term $-rU^2$. Finally, the term $U'\left\langle uv\right\rangle$ can be interpreted as the energy injection rate in the zonal component of the flow.
On the other hand, equation (\ref{eq:formule1 dim}) can
be written as 
\begin{equation}
U'\left\langle uv\right\rangle =\epsilon \label{eq:energy transfer}
\end{equation}
after integration over $y$.
From the energy balance (\ref{eq:energy bilan}),
the term $U'\left\langle uv\right\rangle $ can be interpreted as
the rate of energy transferred from the small-scale eddies to the
mean flow. $\epsilon$ is the total energy injection rate. Relation (\ref{eq:energy transfer}) thus means that all
energy injected at small scale is transferred \emph{locally} to the
largest scale of the flow. The fact that all energy is transferred to the largest svale before being dissipated can be explained by the limits $\alpha\rightarrow0$
and $K\rightarrow+\infty$. The inertial limit $\alpha\rightarrow0$
corresponds to a vanishing value of the friction coefficient $r$.
In the limit of vanishing friction, the system has no time to dissipate
energy at small scale and all energy is transferred to the largest scale. The small scale forcing limit $K\rightarrow+\infty$
prevents energy transfers between the different parts of the flow.
The velocity fluctuations at latitude $y$ only interact with the
flow in a small region of size of order $\frac{1}{K}$ around. Thus,
spatial energy transfer is impossible and energy has to be transferred
to the mean flow at the same latitude $y$. For the local velocity fluctuations,
the mean flow at scale $\frac{1}{K}$ looks like a parabolic profile
with derivative $U'(y)$ and second derivative $U''(y)$, that's why
the asymptotic development of the Reynolds stress divergence is expressed
in terms of $U'$ and $U''$. 

To sum up this idea, we can say that the energy transfer is local
in physical space, but nonlocal in Fourier space. Energy is transferred
directly from the scale $\frac{1}{K}$ to the mean flow through direct
interaction between the mean flow and the eddies, and not through
an inverse energy cascade in Fourier space. Energy transfer is possible
only if $U'\neq0$. At the extrema of jets, expression (\ref{eq:energy transfer})
breaks because direct energy transfer from small scales to the mean
flow is impossible.\\

\section{Cusps for eastward jets\label{sub:Cusps-for-eastward}}

\begin{figure}
\begin{centering}
\includegraphics[scale=0.2]{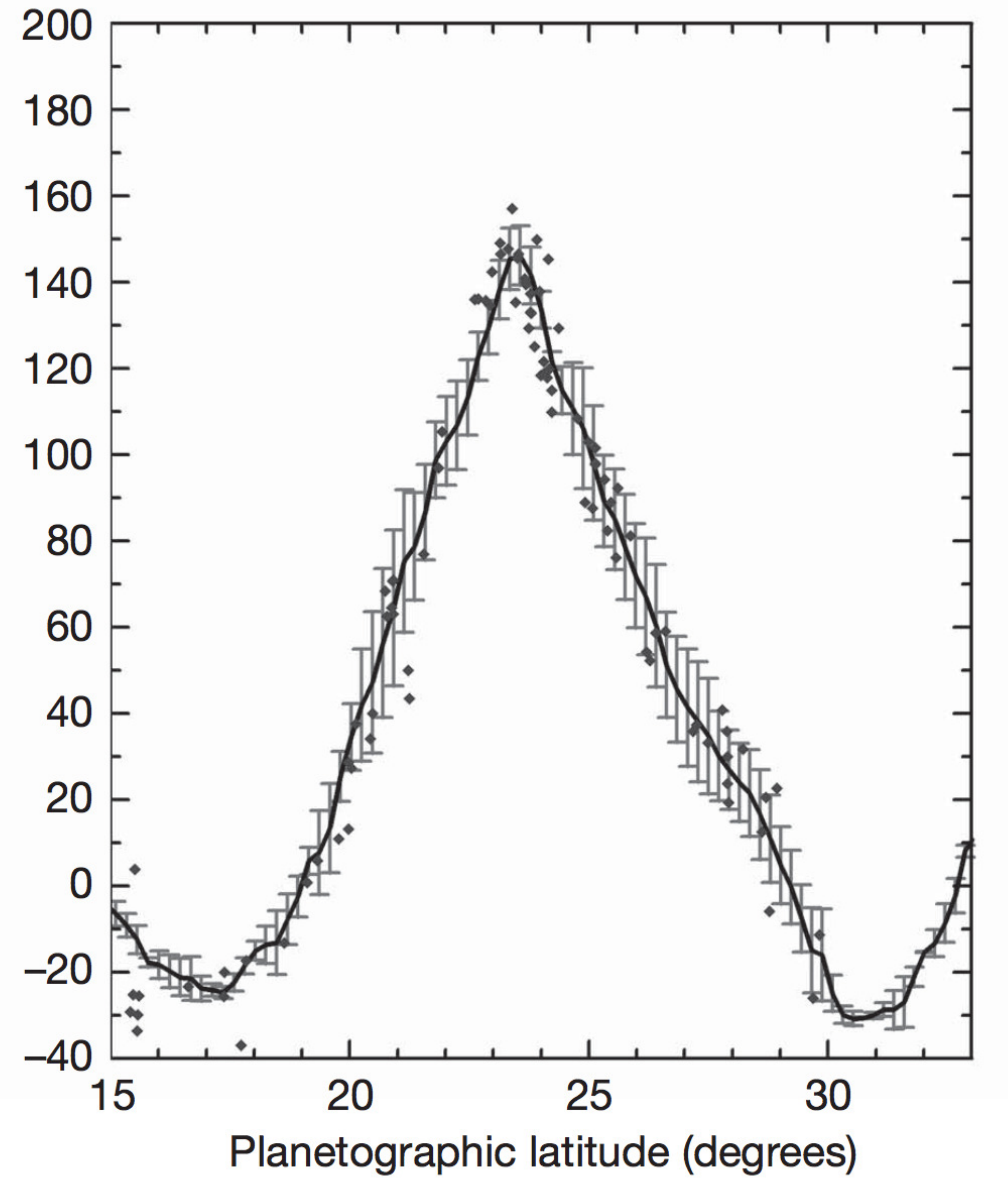} 
\par\end{centering}
\protect\caption{The $24^{o}$N Jupiter eastward jet (taken from \citep{sanchez2008depth}).
The vertical scale is the mean velocity of the wind ($m.s^{-1}$).
\label{fig:The-eastward-jet}}
\end{figure}

We now assume that there are no hydrodynamical instabilities in the eastward part of zonal jets.
In the previous parts of this paper, we saw that the formula (\ref{eq:formule1 dim})
$\partial_{y}\left\langle uv\right\rangle =-\epsilon\frac{U''}{U'^{2}}$
gives a divergent mean velocity profile and we discussed that this
formula can be valid only in the limit $\frac{KU'}{U''}\rightarrow\infty$. The latter limit is not satisfied close to the eastward extrema. The result (\ref{eq:particle in potential}) shows that the ratio $\frac{KU'}{U''}$ behaves as $K(y_{cr}-y)$, where $y_{cr}$ is the critical latitude of the eastward divergence. Even for large values of $K$, the asymptotic expansion breaks down in a narrow region of size $\frac{1}{K}$ around the eastward peak.
The limit (\ref{eq:formule1 dim})
is only valid between the extrema of the jet. But in a region of size $\frac{1}{K}$
around the extremum, another mechanism takes place to stop the
jet growth, and regularize the mean velocity profile at scale $\frac{1}{K}$. On Jupiter, the data collected by Gallileo and Cassini
probes, displayed in figure~(\ref{fig:The-eastward-jet}), indicate
that the eastward jets have ``cusps'', while westward jets seem
smoother. We first discuss eastward jet cusps.

Looking more precisely on the cusp of figure~(\ref{fig:The-eastward-jet}),
we see that its size is approximately 1 degree i.e a scale of about $1000$
km. When we observe Jupiter's surface, we can see the fluctuating
vortices evolving in a timescale of a few days (\citep{porco2003cassini}). The size of those vortices
are related to three dimensional motions, producing convection plumes,
that develop potential vorticity disturbances at a scale which approximately
the Rossby deformation radius of order $1,000\,\mbox{km}$ and with
potential vorticity of order $\beta$, the Coriolis parameter. In
our effective model of barotropic flows, all these convective phenomena
are modeled by the stochastic force. Accordingly, we choose the forcing scale $\frac{1}{K}$ to be  of the order of a thousand kilometers.

A natural question is: can we have a cusp solution of the stationary equation
\[
\left\langle v\omega\right\rangle [U]=U,
\]
in the limit $K\rightarrow\infty$?

In order to adress this question, we consider equation (\ref{eq:inhomogeneous rayleigh}) and study its large $K$ asymptotic after changing the scale $y\leftarrow Ky$.
We denote $\theta$ the angle defined through the relation $\cos\theta:=\frac{k}{K}$.
As we are looking for a cusp of size $\frac{1}{K}$, it will be convenient
to set $\tilde{U}(y)=U\left(\frac{y}{K}\right)$. This implies that
$\frac{1}{K^{2}}U''\left(\frac{y}{K}\right)=\tilde{U}"(y)$. Equation
(\ref{eq:inhomogeneous rayleigh}) becomes
\[
\left(\frac{d^{2}}{dy^{2}}-\cos^{2}\theta\right)\varphi_{\epsilon}(y/K,c)+\frac{\beta/K^{2}-\tilde{U}"(y)}{\tilde{U}(y)-c-i\epsilon}\varphi_{\epsilon}(y/K,c)=\frac{\omega(y/K,0)}{ikK^{2}(\tilde{U}(y)-c-i\epsilon)}.
\]
We set $ikK^{2}\varphi(y/K,c):=\phi(y,c)$. From (\ref{eq:omegainfini}) the function $\omega^{\infty}$
satisfies 
\begin{equation}
\omega^{\infty}\left(\frac{y}{K}\right)=\omega\left(\frac{y}{K},0\right)+\left(\tilde{U}"(y)-\frac{\beta}{K^{2}}\right)\phi_{+}(y,c).\label{eq:omega_inf_Klim}
\end{equation}
Expression (\ref{eq:omega_inf_Klim}) shows that the limit of large
$K$ completely cancels the effect of the parameter $\beta$. However
the solution $\tilde{U}$ still depends on $\theta$. We first consider
the case where the spectrum has only one component $\theta$ . Let
$w_{\theta}(y):=\omega^{\infty}\left(\frac{y}{K}\right)$ . Equations (\ref{eq:inhomogeneous rayleigh}),(\ref{eq:omegainfini}) and (\ref{eq:inertial}) give the set
of equations defining the Reynolds stress divergence $\left\langle v\omega\right\rangle [\tilde{U]}$ in the large $K$ limit
\begin{eqnarray}
\left(\frac{d^{2}}{dy^{2}}-\tan^{2}\theta\right)\phi_{\epsilon}(y,c)-\frac{\tilde{U}"(y)}{\tilde{U}(y)-c-i\epsilon}\phi_{\epsilon}(y,c) & = & \frac{e^{i\sin\theta y}}{\tilde{U}(y)-c-i\epsilon}\nonumber \\
e^{i\sin\theta y}+\tilde{U}"(y)\phi_{+}(y,\tilde{U}(y)) & = & w_{\theta}(y)\nonumber \\
\frac{1}{\tilde{U}"(y)}[|w_{\theta}(y)|^{2}-1] & = & \left\langle v\omega\right\rangle _{\theta}[\tilde{U]}.\label{eq:system-cusp}
\end{eqnarray}
The first equation is the inhomogenous Rayleigh equation without $\beta$
effect. The second one is the modified expression to compute $\omega^{\infty}$,
and the last one is the pseudomomentum balance giving access to the
Reynolds stress divergence.

Before we go on with numerical analysis, let us give some analytic
results on the set of equations (\ref{eq:system-cusp}). 
\begin{itemize}
\item We have already given expression (\ref{eq:formule1 adim}) for the
Reynolds stress divergence in the limit $K\rightarrow+\infty$, away
from the extremum of the jet. As we have used the scaling $y\leftarrow Ky$ to find (\ref{eq:system-cusp}),
we expect to recover the asymptotic (\ref{eq:formule1 adim}) in
the limit $y\rightarrow\infty$. For a given profile $\tilde{U}$,
we have the asymptotic result $\left\langle v\omega\right\rangle [\tilde{U}]\underset{y\rightarrow\infty}{\sim}-\frac{\tilde{U}"}{\tilde{U}'^{2}}.$ 
\item We know that the relation $\omega^{\infty}(y_{cr})=0$ holds at the
extremum (see subsection \ref{sub:Computation-of}), which corresponds
here to $w_{\theta}(y_{cr})=0$. At the extremum, the third equality
in (\ref{eq:system-cusp}) shows that $\left\langle v\omega\right\rangle =-\frac{1}{\tilde{U}"}$.
At a maximum of $U$ (eastward jet), $\tilde{U}"<0$ and the Reynolds
stress divergence thus forces the profile $\tilde{U}$ to grow. The contrary
happens at a minimum of $U$: we have $\tilde{U}">0$ and the velocity decays, such that
its magnitude grows. The consequence is that the turbulence always forces
the jet to grow. The growth can  be stopped by either
linear friction or non linear effects beyond the quasilinear approximation.
For westward jets, we will see in section (\ref{sub:The-question-of})
that it can also be stopped by an hydrodynamic instability. 
\item The formula 
\begin{equation}
\left\langle v\omega\right\rangle (y_{cr})=-\frac{1}{\tilde{U}"(y_{cr})}\label{eq:relation_cusp1}
\end{equation}
is in itself noteworthy. It comes from the phenomenon of depletion
of vorticity at the stationary streamlines, which has been already
emphasized by \citep{Bouchet_Morita_2010PhyD}. To reach the stationary
profile, $\left\langle v\omega\right\rangle $ has to equilibrate
the linear friction. At the jet extremum, the stationary state
of $U$ in (\ref{eq:quasilinear})  gives the equality 
\begin{equation}
\left\langle v\omega\right\rangle (y_{cr})=\tilde{U}(y_{cr}).\label{eq:relation_cusp2}
\end{equation}
We can thus link the value of the velocity at the extremum of the
jet and the curvature of the cusp. Relations (\ref{eq:relation_cusp1}-\ref{eq:relation_cusp2})
give the second important result of this paper
\begin{equation}
\tilde{U}(y_{cr})=-\frac{1}{\tilde{U}''(y_{cr})}.\label{eq:curvature adim}
\end{equation} Coming back to dimensional
fields, the eastward cusp satisfies the relation
\begin{equation}
U(y_{cr})=-\frac{\epsilon K^{2}}{rU''(y_{cr})},\label{eq:curvature}
\end{equation}
where $\epsilon$ is the energy injection rate. Relation (\ref{eq:curvature})
is a universal property of stationary jet profiles. It relates the
strength of a jet to its curvature, and the physical parameters $\epsilon,r$
and $K$. It does not depends on the forcing Fourier spectrum, but
only on the scale $\frac{1}{K}$ at which energy is injected.\\
\end{itemize}
\begin{figure}
\begin{centering}
\includegraphics[scale=0.2]{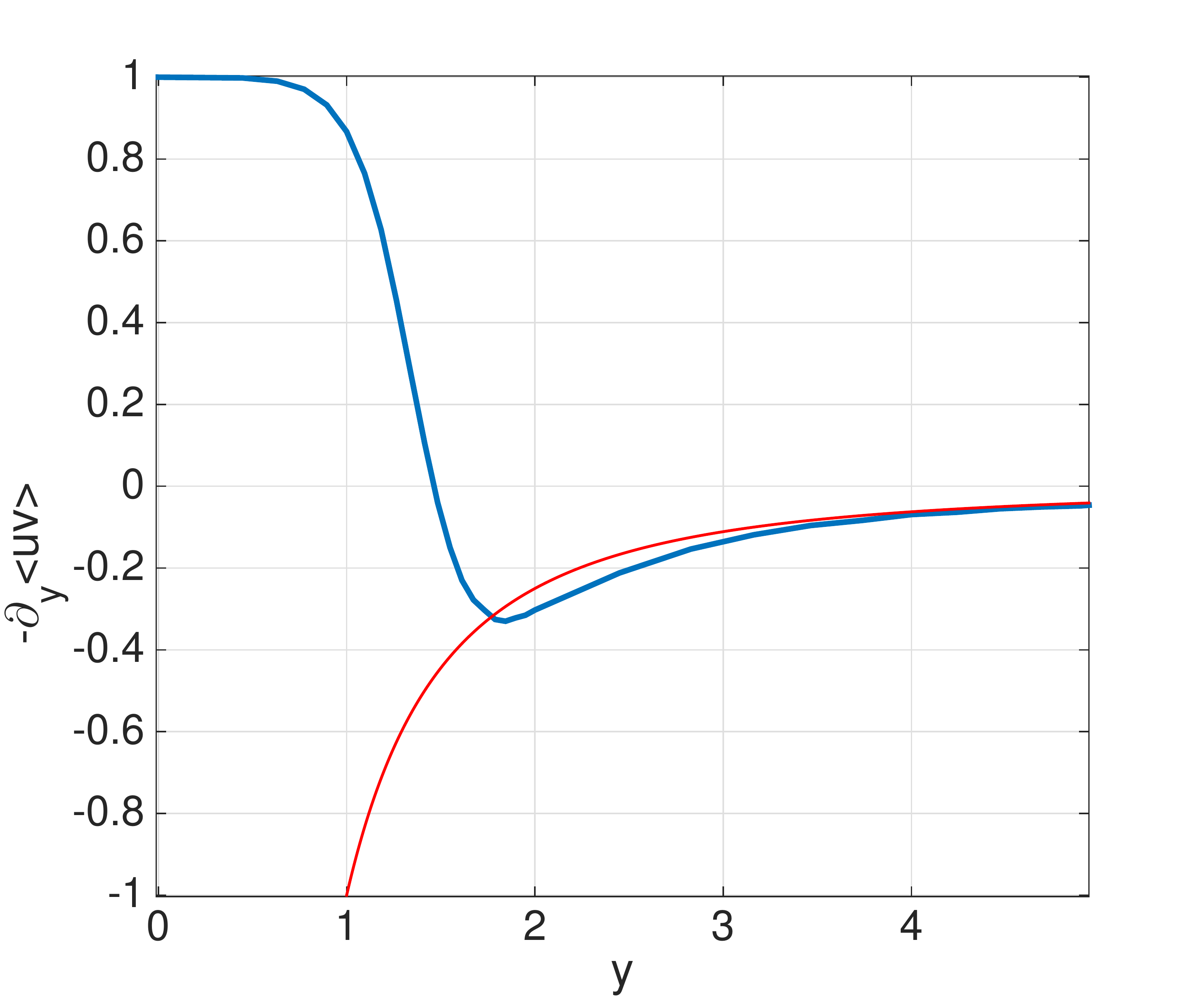}
\includegraphics[scale=0.2]{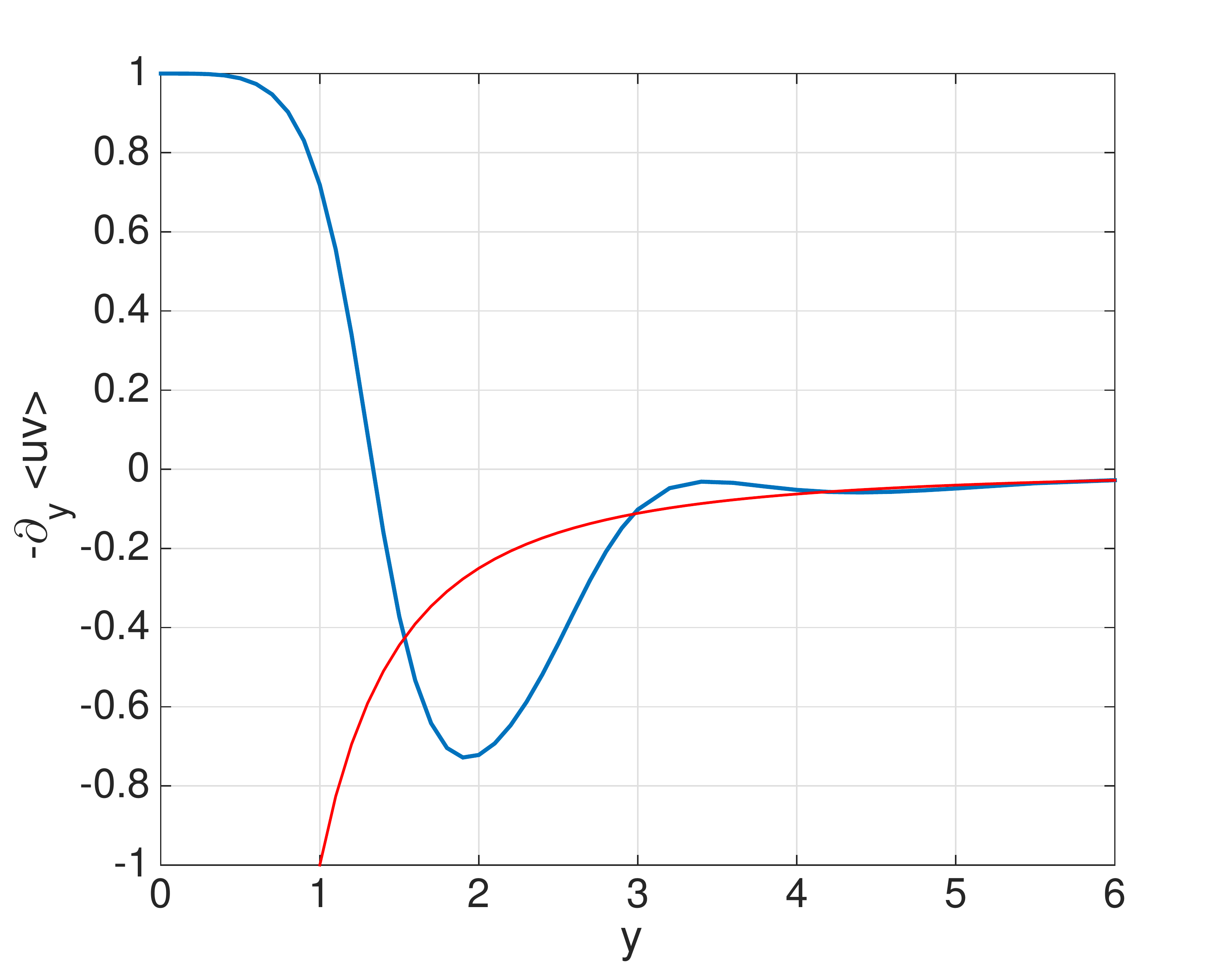}
\par\end{centering}
\protect\caption{Reynolds stresses close to an eastward jet.
Left: the Reynolds stress divergence $-\partial_{y}\left\langle uv\right\rangle $
from (\ref{eq:system-cusp}) for a parabolic profile $\tilde{U}(y)=-\frac{y^{2}}{2}$,
and\textbf{ $\theta=\frac{\pi}{8}$}. (thick curve). The thin curve
depicts the theoretical asymptote $\frac{\tilde{U}"}{\tilde{U}'^{2}}$.
As one can notice, $-\partial_{y}\left\langle uv\right\rangle (0)=-\frac{1}{\tilde{U}"}=1$ in agreement with the theoretical result (\ref{eq:relation_cusp1}). Right: the total Reynolds
stress divergence resulting from the sum over $\theta\in\left[-\frac{\pi}{3};\frac{\pi}{3}\right]$.
The Reynolds stress divergence has no universal expression in the
intermediate region between the cusp (at $y=0$) and the asymptotic
region $y\gg1$. The cusp velocity profile has thus to be computed
numerically using the particular shape of the force Fourier spectrum.
\label{fig:quadratic extremum}}
\end{figure}

Let us illustrate those results by a numerical computation of the
nondimensional equations (\ref{eq:system-cusp}). The numerical computation goes the following way: we solve the first
equation of (\ref{eq:system-cusp}), for given values of the Laplace transform parameter $\epsilon$ and the phase velocity $c$, and given boundary conditions.
We impose vanishing boundary conditions at infinity for $\phi_{\epsilon}$.
$\epsilon$ has to be small because we want to compute the solution
$\phi_{+}$ when $\epsilon$ goes to zero. The left panel of figure~(\ref{fig:quadratic extremum})
has been obtained with $\epsilon=10^{-5}$. Because the solution $\phi_{+}$
has a singularity at $U(y_{c})=c$, an extreme precision is required
to obtain convergence of the numerical calculations. To obtain
the value of $\left\langle v\omega\right\rangle _{\theta}(y)$, we
have to compute the solution $\phi_{+}$ for $c=\frac{y^{2}}{2}$,
and this has to be done for each value of $y$. On the left of figure~(\ref{fig:quadratic extremum}),
about 20 values of $y$ were used to plot the blue curve.

The plot of the Reynolds stress divergence in figure~(\ref{fig:quadratic extremum})
clearly displays two regions with a sharp transition (located around
$y=2$ in the figure). In the first region, the mean velocity profile
forms a cusp, which joins continuously the second region of large $y$
values. The second region corresponds to the domain where the expression
$\left\langle v\omega\right\rangle =-\frac{U''}{U'^{2}}$ is valid.
The velocity profile joining the cusp to the asymptotic profile of
figure (\ref{fig:diverging profiles}) is non-universal with respect
to the forcing spectrum. In the right panel of figure~(\ref{fig:quadratic extremum}),
we plot $-\partial_{y}\left\langle uv\right\rangle $
for a uniform forcing spectrum in the range $\theta\in\left[-\frac{\pi}{3};\frac{\pi}{3}\right]$.

With the system of equations (\ref{eq:system-cusp}), we have been
able to show that a cusp of typical size $\frac{1}{K}$ forms
at the eastward extremum of the jet. This cusp regularizes the velocity
profile at its maximum and stops the divergence observed in figure
(\ref{fig:diverging profiles}). The relation between the curvature
of the jet at the extremum and its maximal velocity (\ref{eq:curvature})
is universal as it does not involve the explicit expression of the
spectrum of the stochastic force. However, the exact velocity profile
joining the cusp to the asymptotic profile of figure (\ref{fig:diverging profiles})
is rather complicated and is not at all universal.

\section{Computation of Reynolds stress divergence for westward jets\label{sub:Computation-of-Reynold's}}

As explained in section \ref{sub:Cusps-for-eastward}, the parameter
$\beta$ disappears from the equations when we try to compute the
equilibrium profile in the small scale forcing limit $K\rightarrow\infty$,
because the $\beta$ effect becomes irrelevant at the scale $\frac{1}{K}$.
Using this approach, we could expect the jet to be symmetric with
respect to the transformation $U\rightarrow-U$. At a formal level,
nothing in equations (\ref{eq:close equation velocity}) nor (\ref{eq:system-cusp})
seems to make any difference between the eastward and
the westward part of a jet. However, a look at the jets observed on Jupiter
shows a clear asymmetry between eastward and westward jets, especially
at high latitudes. One key point is that, as clearly stated, the previous
sections assume that the linearized equations close to the jet are
stable, and do not sustain neutral modes.

\begin{figure}
\begin{tabular}{cc}
\includegraphics[scale=0.2]{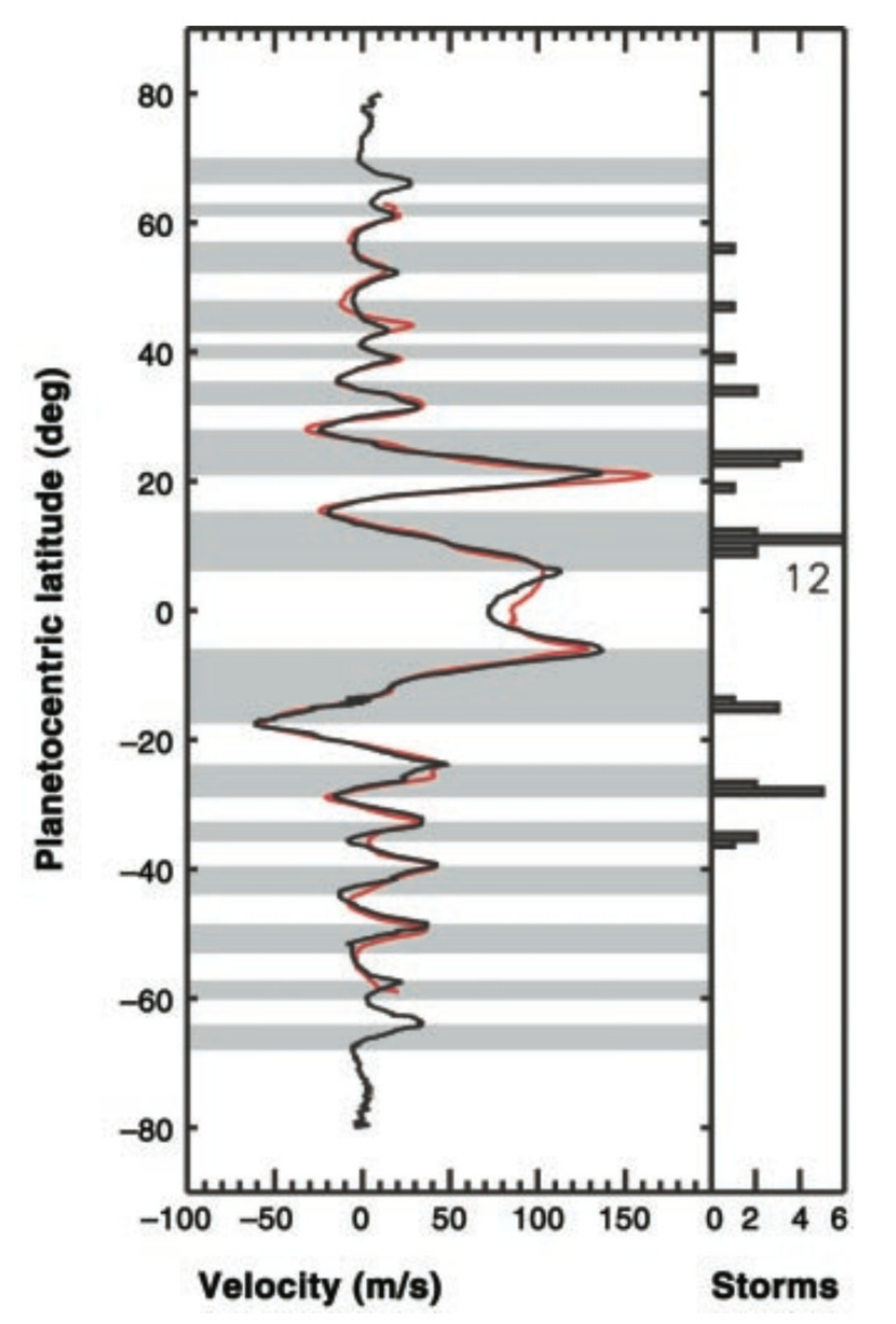}  & \includegraphics[scale=0.29]{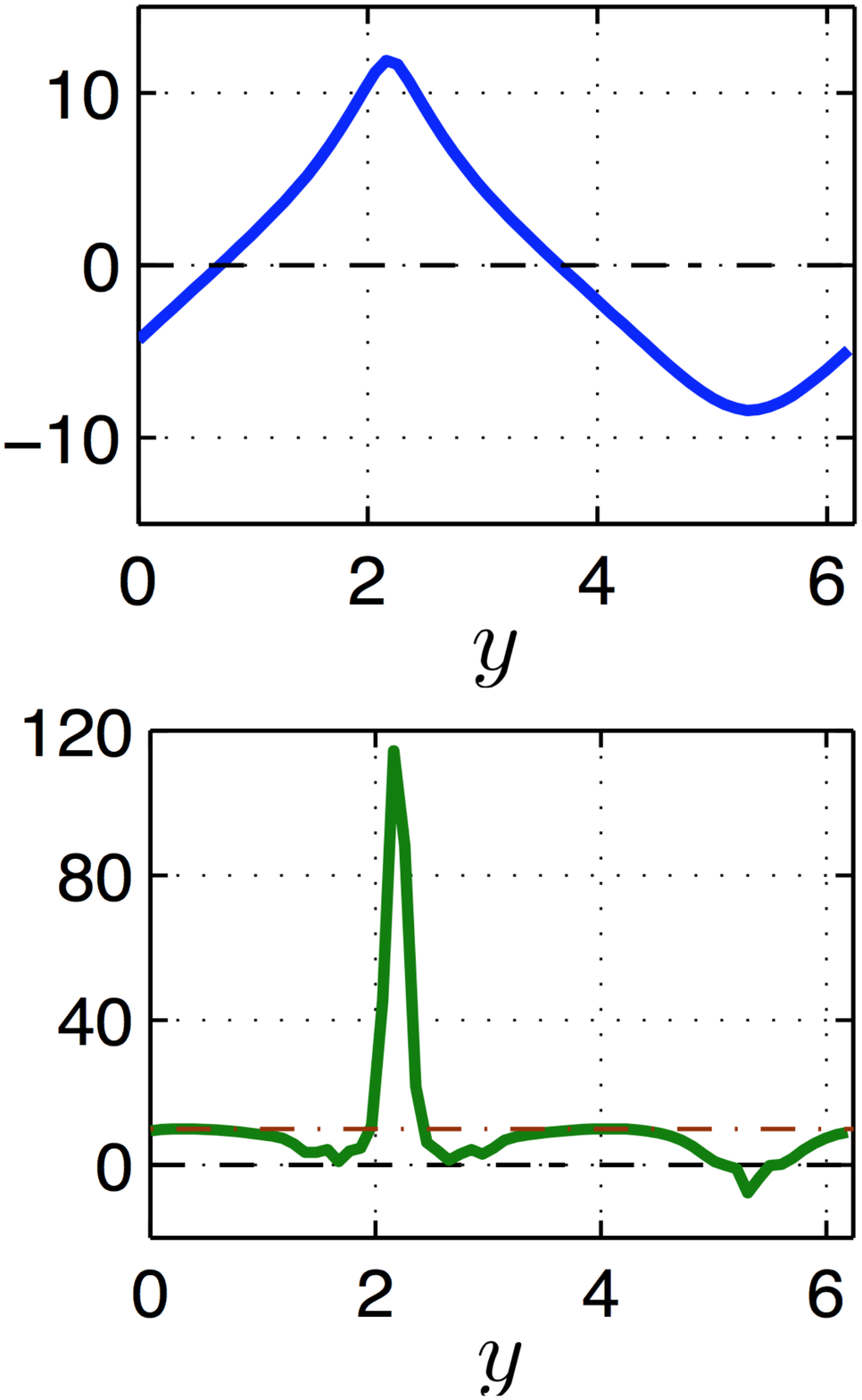} \tabularnewline
\end{tabular}

\protect\caption{Left panel: zonal jets on Jupiter. Data collected by the Gallileo
and Cassini probes (from \citep{porco2003cassini}). Right panel: A numerical
simulation of the quasilinear barotropic equations (S3T system) performed
in \citep{constantinou2015formation}.The top figure displays the
mean velocity profile $U$ and the bottom figure displays $\beta-U''$.
The cusp is obvious on both figures, the peak on the bottom figure corresponds to the eastward
extremum of the jet. $\beta-U''$ is always positive, thus satisfying
the Rayleigh-Kuo criterion except possibly at the westward extremum.\label{fig: jupiter jets}}
\end{figure}

On Jupiter's jets, cusps only exist on the eastward part whereas the
westward part looks like a parabolic profile with curvature between
$2\beta$ and $3\beta$ (see figure~(\ref{fig: jupiter jets}) and
\citep{ingersoll1981interaction} for a discussion on the value of
the curvature). Numerical simulations of the barotropic model also
show this asymmetry.
In \citep{constantinou2015formation} for example, the curvature at
the eastward jet is almost exactly $\beta$ and seems to be trapped
at this value whatever large the coefficients $K$ and $\frac{1}{\alpha}$
are. The value of $\beta-U''$ is always positive, and the Rayleigh-Kuo
criterion for jet stability is satisfied. The aim of this section
is to understand what is the behavior of a parabolic jet with $U''$
close to $\beta$ and see if the profile $\beta\frac{y^{2}}{2}$ can
or not be a stationary solution of the barotropic model (\ref{eq:Navier-stokes adim}).

\subsection{Modified Rossby waves}

We consider in equation (\ref{eq:inhomogeneous rayleigh}) a parabolic
profile $U(y)=\gamma\frac{y^{2}}{2}$, and we want to study the behavior
of the Reynolds stress divergence when $\gamma$ is close to $\beta$.
For $\gamma=\beta$, any perturbation is carried freely by the mean
flow, and equation (\ref{eq:deterministic}) reduces to 
\[
\partial_{t}\omega+ikU\omega=0,
\]
which is easily solved by $\omega(y,t)=\omega(y,0)e^{-ikUt}$. Expression
(\ref{eq:inertial}) is then singular, because $U''-\beta$ vanishes in
the denominator, and $|\omega^{\infty}|^{2}-1=|\omega(y,0)|^{2}-1=0.$
If we try to compute directly the Reynolds stress divergence $\left\langle v\omega\right\rangle $,
we will find a singularity in $y=0$. Therefore, the aim is to compute
$\left\langle v\omega\right\rangle $ for $\gamma$ smaller and larger
than $\beta$ and let then $\gamma\rightarrow\beta$.

It has been proved long ago that for $0<\gamma<\beta$ we
have modified Rossby waves in the flow \citep{drazin1982rossby},
with at least one Rossby wave as soon as $\gamma<\beta$. In \citep{brunet1990dynamique},
the case of a parabolic profile is thoroughly studied and a method
is found to compute the Rossby waves and their velocity. Basically,
it consists in doing a Fourier transform in $y$ and transform the
Rayleigh equation into a one dimensional Schr{\"o}dinger equation.
The one dimensional Schr{\"o}dinger equation describes a particle
in a potential vanishing at infinity. Possible bound states of the
Schr{\"o}dinger equation correspond to modified Rosby waves.

For $\gamma>\beta$, the Schr{\"o}dinger equation potential is positive,
and classical results prove that there is no bound state, and thus
there is no Rossby waves. In that case expression (\ref{eq:inertial})
will be valid to compute the Reynolds stress divergence. By contrast,
for $0<\gamma<\beta$, the Sch{\"o}dinger equation potential is negative
(the position zero is attractive). Classical results \citep{reed1978modern}
shows that there exists a least one bound state. There is thus at
least one modified Rossby wave. Moreover as the potential deepens
for decreasing $\gamma/\beta$, the number of bound states and thus
the number of modified Rossby waves increases when $\gamma/\beta$
decreases. When $0<\gamma<\beta,$ because of the presence of waves,
we have to use expression (\ref{eq:inertialmodes}) to compute the
Reynolds stress divergence.

We discuss more precisely the existence of Rossby waves and their computation for a parabolic profile  in appendix
\ref{sec:Modified-Rossby-waves}.

\subsection{Singularity of the Reynolds stress for a jet curvature close to $\beta$ \label{sub:Singularity-of-the}}

We now compute the Reynolds stress divergence $\left\langle v\omega\right\rangle $ using
the same method as for the cusp case discussed in section \ref{sub:Cusps-for-eastward},
but without taking the limit $K\rightarrow\infty$. It happens that
the parabolic profile has an additional symmetry, it is invariant
under the transformation $y\leftarrow Ky$. For a parabolic profile,
equation (\ref{eq:inhomogeneous rayleigh}) only depends on the parameter $\tan\theta:=\frac{l}{k}$
and $\mu:=1-\frac{\beta}{\gamma}$. As discussed previously, Rossby
waves appear when $\mu<0$ (equivalently for $0<\gamma<\beta$).

 Using equations (\ref{eq:inhomogeneous rayleigh}),(\ref{eq:omegainfini}) and (\ref{eq:inertialmodes}), the self-consistent equations for the jet write 
\begin{eqnarray}
\left(\frac{d^{2}}{dy^{2}}-\tan^{2}\theta\right)\varphi_{\epsilon}(y,c)-\frac{\mu}{\frac{y^{2}}{2}-c-i\epsilon}\varphi_{\epsilon}(y,c) & = & \frac{\mathcal{P}e^{i\sin\theta y}}{\frac{y^{2}}{2}-c-i\epsilon}\nonumber \\
\mathcal{P}e^{i\sin\theta y}+\mu\varphi_{+}\left(y,\frac{y^{2}}{2}\right) & = & w_{\theta}(y)\nonumber \\
\frac{1}{\mu}\left[\left|\left(1-\mathcal{P}\right)e^{i\sin\theta y}\right|^{2}+\left|w_{\theta}(y)\right|^{2}-1\right] & = & \frac{1}{\gamma}\left\langle v\omega\right\rangle _{\theta}[\tilde{U]}.\label{eq:system-cusp-1}
\end{eqnarray}
We have denoted by $\mathcal{P}$ the projector on the space orthogonal
to the neutral modes.

\begin{figure}
\includegraphics[scale=0.2]{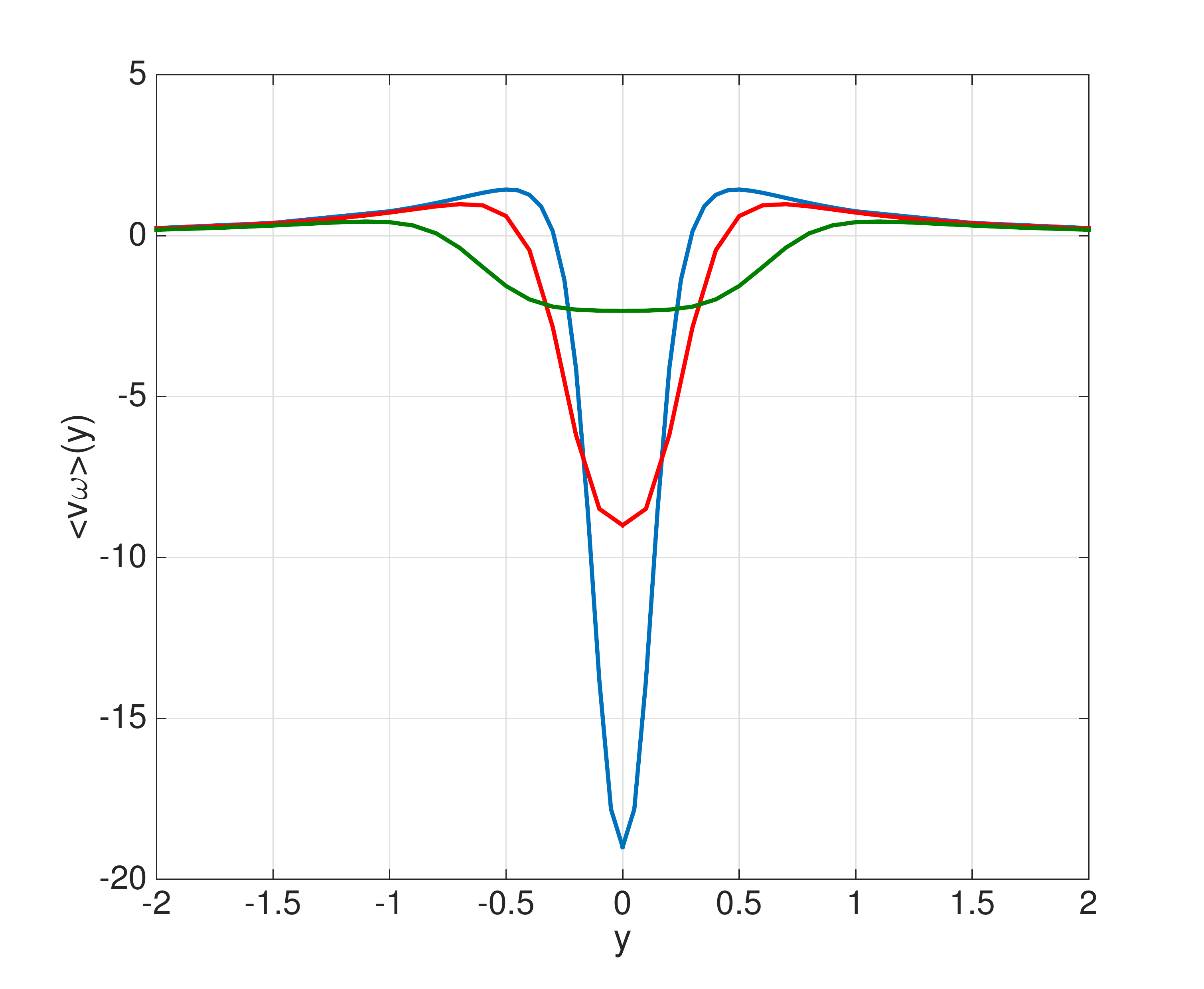}
\includegraphics[scale=0.2]{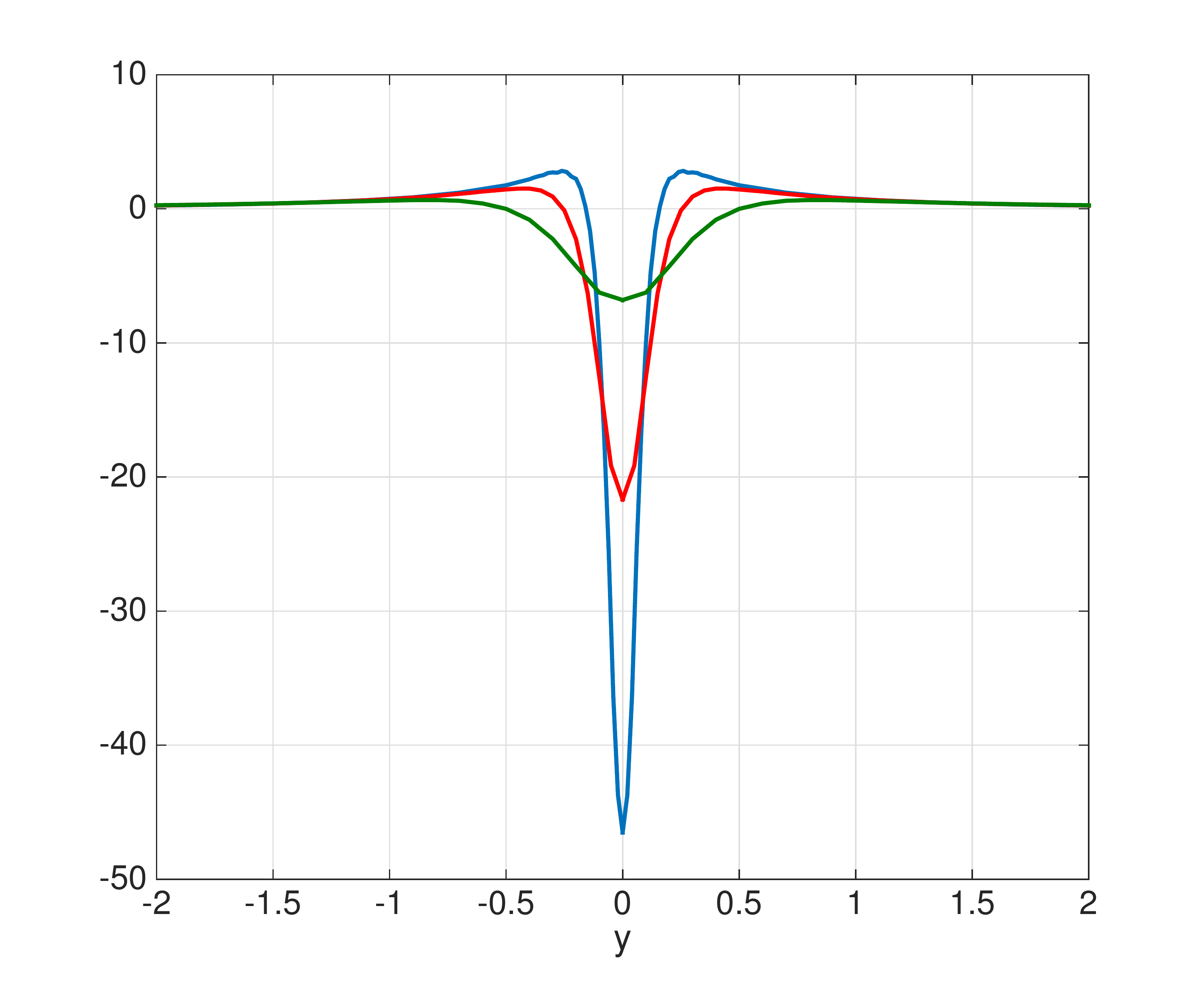}

\protect\caption{Reynolds stresses  close to a westward jet.
Left: Minus the Reynolds stress divergence $<v\omega>$ obtained for
$\mu=0.3$ (green), $\mu=0.1$ (red) and $\mu=0.05$ (blue). When
the parameter $\mu$ comes closer to zero, the divergence at $y=0$
becomes more and more pronounced. Right: Minus the Reynolds stress divergence
for $\mu=-0.3$ (green), $\mu=-0.1$ (red) and $\mu=-0.05$ (blue).
Other parameters are $\theta=\frac{\pi}{8}$ and $\beta=1$.\label{fig:Reynold for parabola}}
\end{figure}

The result of the numerical integration of (\ref{eq:system-cusp-1})
is shown in figure~(\ref{fig:Reynold for parabola}). The main result
is that the stress $\left\langle v\omega\right\rangle $ has the same
qualitative behavior both for $\gamma<\beta$ and for $\gamma>\beta$.
For $\gamma>\beta$, i.e $\mu>0$, we still have the result that $w_{\theta}(0)=0$,
which implies that the stress $\left\langle v\omega\right\rangle $
is diverging as $-\frac{1}{\mu}$ when $\mu\rightarrow0^{+}$. For
$0<\gamma<\beta$, i.e $\mu<0$, the stress is diverging as $\frac{[|\omega_{\theta}^{1}(0)|^{2}-1]}{\mu}$
when $\mu\rightarrow0^{-}$. $\omega_{\theta}^{1}$ is the projection
of $e^{i\sin y}$ on the first neutral mode. It happens that $|\omega_{\theta}^{1}(0)|^{2}-1$
is always positive. Hence, $\frac{[|\omega_{\theta}^{1}(0)|^{2}-1]}{\mu}$
is negative. We conclude that whatever the sign of $\mu$, the stress
$\left\langle v\omega\right\rangle $ has a negative divergence at
the minimum of the jet that makes the jet grow. 

If the curvature $\gamma$ is smaller than $\beta$, the effect of
the Reynolds stress divergence is to narrow the jet and increase the
curvature. When $\gamma$ becomes larger than $\beta$, the quasilinear theory predicts that the jet should continue its growth, and form a cusp exactly the same way as for the eastward jet.
No mechanism in the quasilinear dynamics can stop the growth of the westward jet. To explain the
numerical simulations, we thus have to consider other hypothesis than
the ones considered so far. Among those, we have assumed there is no hydrodynamic instability
in the set of equations (\ref{eq:system-cusp-1}), i.e a mode with
nonzero imaginary part of the velocity. With $\gamma>\beta$ the Rayleigh--Kuo criterion
is violated, the stability of a jet is no longer guaranteed. In the last section of this paper, we will study qualitatively
the effect of an instability to see whether it can really stop the
growth of the westward jet.

\subsection{Hydrodynamic instability in the westward jet\label{sub:The-question-of}}

An unstable mode is a solution of the
homogeneous Rayleigh equation 
\begin{equation}
\left(\frac{\partial^{2}}{\partial y^{2}}-k^{2}\right)\psi+\frac{\beta-U''}{U-c}\psi=0\label{eq:Rayleigh homogene}
\end{equation}
with \emph{complex} phase speed $c$. In particular for unstable modes,
the imaginary part satisfies $kc_{i}>0$ and the consequence
is the exponential growth of a disturbance $\left|\psi(y,t)\right|\propto e^{kc_{i}t}$.
An unstable mode has a contribution to the Reynolds stress divergence. In stationary state, the flow can only sustain unstable modes satisfying $kc_{i}<\alpha$, otherwise the exponential growth of the unstable mode would create a divergence in the Reynolds stress. If the flow sustain unstable modes, we have to modify expression (\ref{eq:inertialmodes}) taking into account the presence of unstable modes.
 We do not report the computation, it is similar to
the one developed in appendix \ref{sec:The-Reynold's-stress} for
neutral modes. Please note that by contrast to neutral modes, the real
part $c_{r}$ of the (complex) speed $c$ lies within the range of $U$ (\citep{drazin2004hydrodynamic,drazin1982rossby}).
The contribution of
an instability in the Reynolds stress has been already computed in
the deterministic case (\citep{PedloskyBook} p 576), and we modify
here the classical result to adapt it to the stochastic case.

Let $\omega^{c}(y)$ be the projection of the initial condition $e^{ily}$
on the unstable mode, and $\psi^{c}$ the associated stream function
defined by $\left(\frac{\partial^{2}}{\partial y^{2}}-k^{2}\right)\psi^{c}=\omega^{c}$.
The projection refers to the scalar product induced by the pseudomomentum
conservation law (see appendix \ref{sec:The-Reynold's-stress} for
the discussion). Then the dominant contribution of the unstable mode in the computation of $2\alpha\left\langle |\omega|^{2}\right\rangle $
writes 
\begin{eqnarray}
2\alpha\left\langle |\omega|^{2}\right\rangle  & = & 2\alpha\int_{-\infty}^{0}{\rm d}t~e^{2\alpha t}\left|\omega^{c}(y)e^{ikct}\right|^{2}\\
 & = & |\omega^{c}(y)|^{2}\frac{2\alpha}{2\alpha-2kc_{i}}\label{eq:resultat5-1}.
\end{eqnarray}
Equation (\ref{eq:Rayleigh homogene}) shows that 
$$
\omega^{c}=-\frac{\beta-U''}{U-c}\psi^{c}.\label{eq:resultat5-2}
$$
Equations (\ref{eq:resultat5-1}) and (\ref{eq:resultat5-2}) give the main contribution of an hydrodynamic instability to the Reynolds stress divergence
\begin{equation}
\frac{2\alpha\left\langle |\omega|^{2}\right\rangle }{U''-\beta}=-\frac{2\alpha}{2\alpha-2kc_{i}}\frac{|\psi^{c}|^{2}}{|U-c|^{2}}(\beta-U'').\label{eq:contribution instable}
\end{equation}
Let us emphasize once more that this term is a \emph{contribution}
to the Reynolds stress adding to the other terms coming from the effect
of neutral modes and from $\omega^{\infty}$. The important
point in (\ref{eq:contribution instable}) is that the coefficient
$\frac{2\alpha}{2\alpha-2kc_{i}}\frac{|\psi^{c}|^{2}}{|U-c|^{2}}$
is strictly positive, which means that the term coming from the unstable
mode \emph{opposes} a change of sign of $\beta-U''$. In order to equilibrate, the
jet needs to make a continuous barotropic adjustment of the mean flow curvature at the westward edge.\\

\begin{figure}
\includegraphics[scale=0.14]{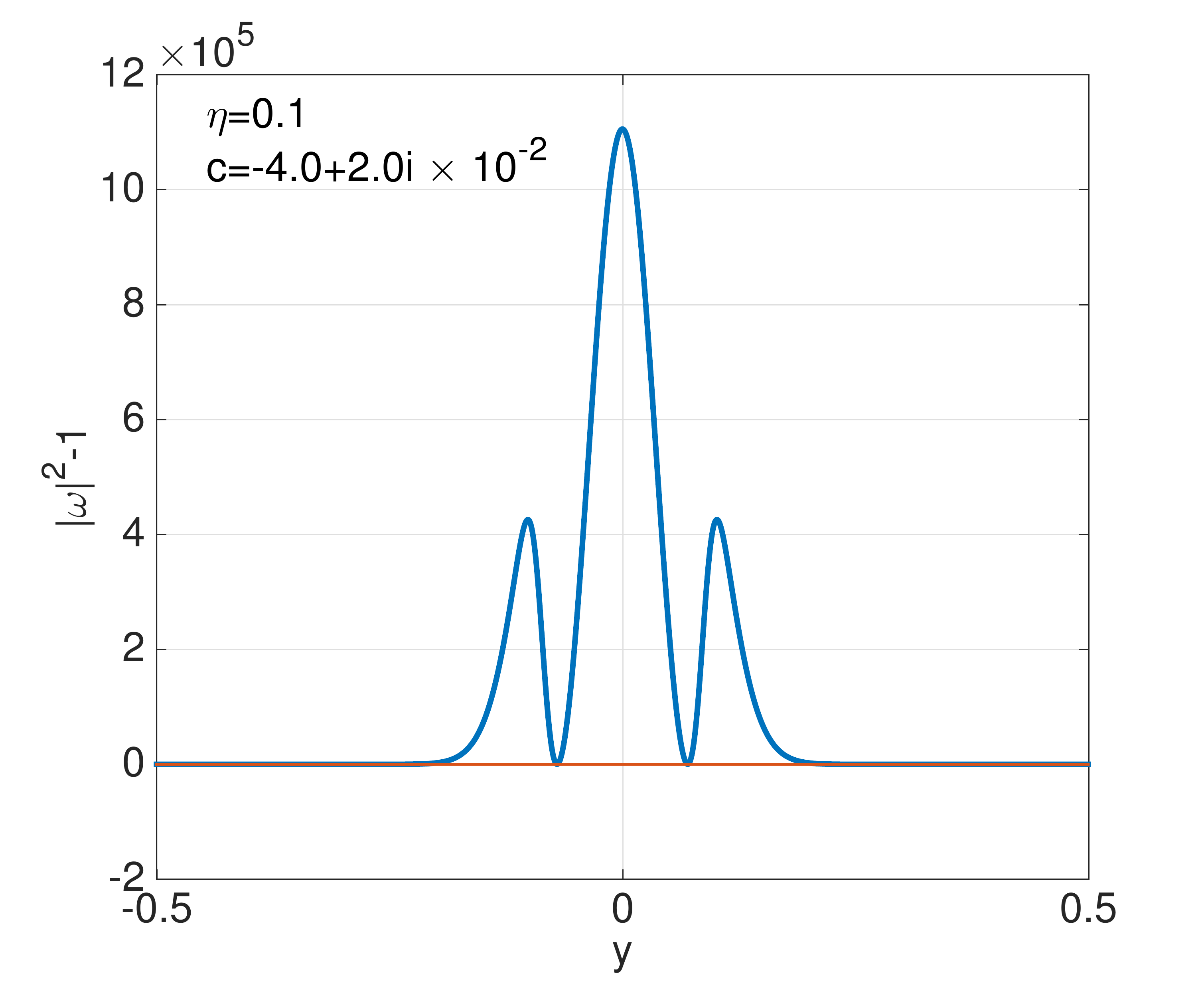}
\includegraphics[scale=0.14]{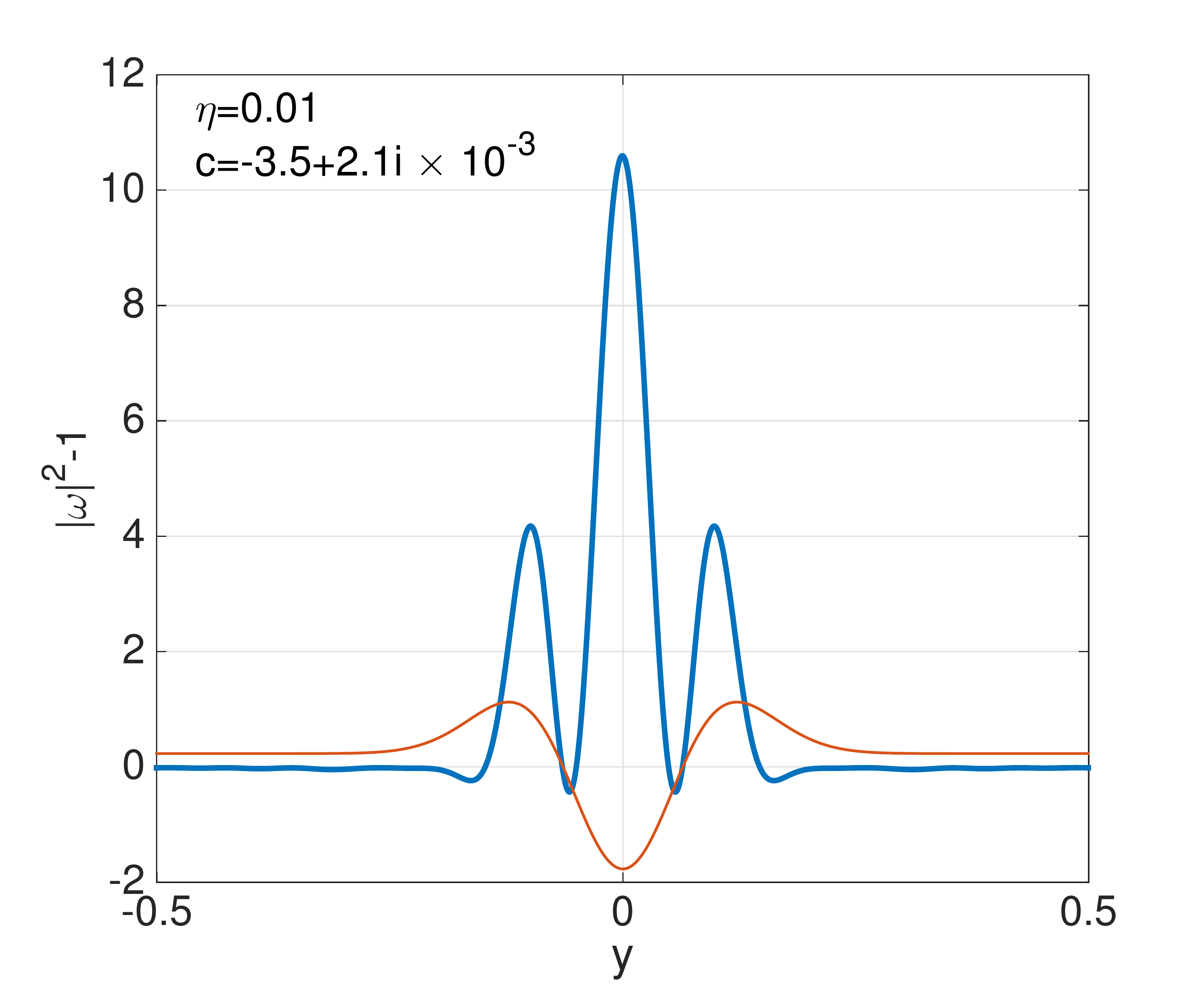}
\includegraphics[scale=0.14]{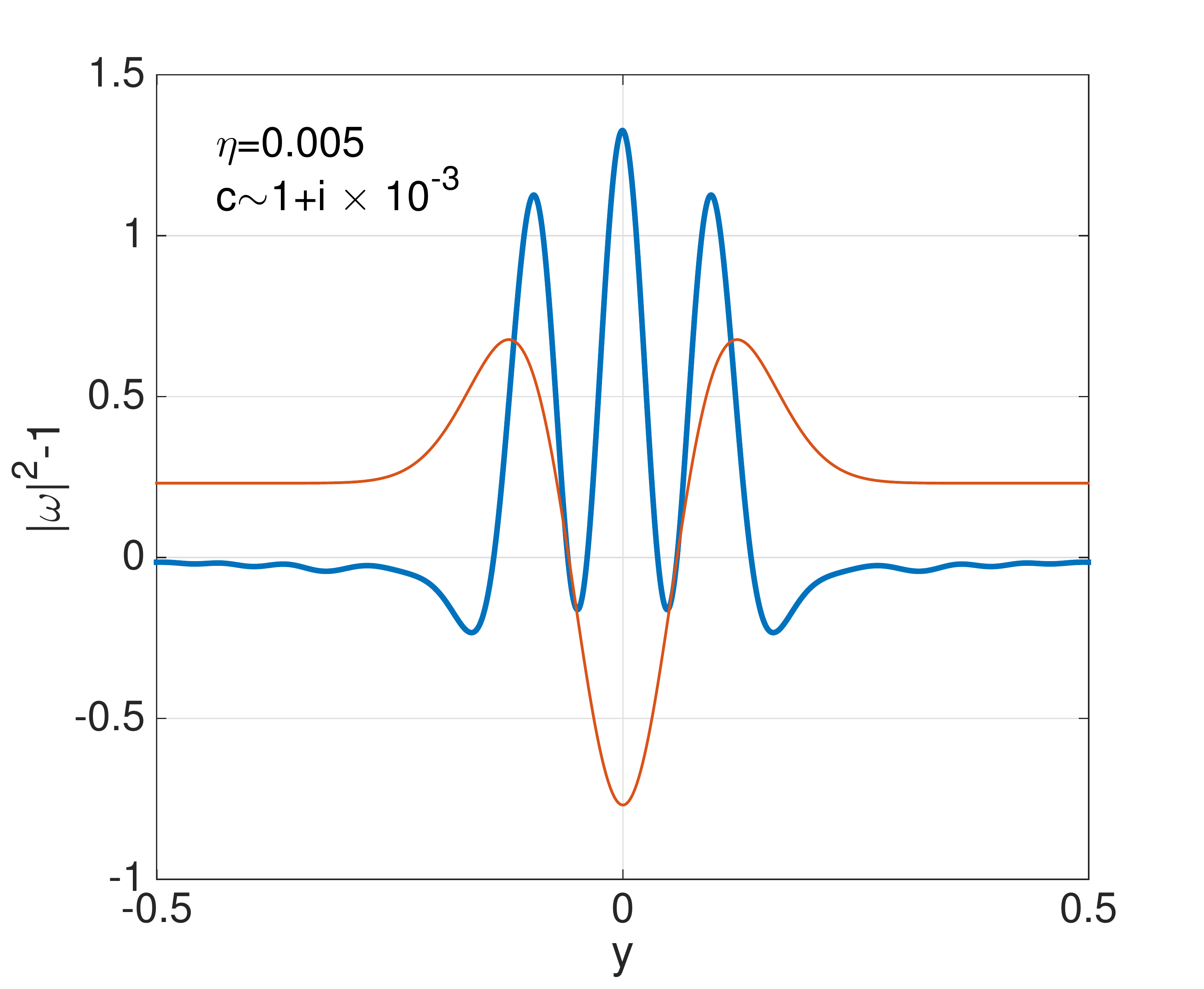}

\protect\caption{Instability growth for a perturbation of an unstable westward jet.
We show here the tensor $|\omega|^{2}-1$ at $T=30$ for different
values of $\eta$, and we compute the value of $c$ whenever possible. The
thiner curve (red color online) shows $\beta-U''$. Close to the instability
threshold, it is no longer possible to determine the value of $c$
because the instability growth rate is too slow. \label{fig:instabilite}}
\end{figure}

In order to illustrate this assertion, we consider
a configuration where the instability develops. We perform a direct
numerical integration of the equation 
\[
\partial_{t}\omega+ikU\omega+ik(\beta-U'')\psi=0,
\]
using periodic boundary conditions in $y$
and the initial condition $\omega(y,0)=e^{ily}$. We use a Runge-Kutta
algorithm of order 4. The profile $U$ is parabolic with $0<\gamma<\beta$
but we add a small disturbance at the extremum in 0 of the form of
a gaussian $-\eta e^{-\frac{y^{2}}{\sigma^{2}}}$. The disturbance
mimics qualitatively the effect of the forcing described in figure~(\ref{fig:Reynold for parabola}),
it models the fact that the mean velocity profile has a narrow curvature
at its extremum. 
\begin{equation}
U(y)=\gamma\frac{y^{2}}{2}-\eta e^{-\frac{y^{2}}{\sigma^{2}}}.\label{eq:velocity_perturbation}
\end{equation}
The values of the chosen parameters are $\beta=1$, $\mu=1-\frac{\beta}{\gamma}=-0.3$,
$k=l=10$. $\sigma$ quantifies the
width of the disturbance, we chose $\sigma=0.1$. $\eta$ describes the magnitude
of the disturbance and  is the control parameter of the simulation. Results are
displayed in figure~(\ref{fig:instabilite}). The red curve is the
graph of $\beta-U''(y)$. When this quantity is strictly positive
everywhere in the flow, the Rayleigh-Kuo criterion is satisfied and
the flow is stable. With the velocity profile chosen in (\ref{eq:velocity_perturbation}),
the Rayleigh-Kuo criterion is violated around $y=0$ as displayed by  the red
curve in figure~(\ref{fig:instabilite}). The blue curve displays the quantity
$|\omega|^{2}(y,t)-1$ at $T=30$. In our simulations, we clearly
see the three peaks of the blue curve growing exponentially with time,
which indicates the existence of an hydrodynamic instability. From
left to right, we have increased the value of the parameter $\eta$.
The larger $\eta$, the more the Rayleigh-Kuo criterion is violated,
and the faster the instability grows. It has been already emphasized that
$|\omega|^{2}-1$ has to vanish at the same time where $\beta-U''=0$
in the flow, and this is confirmed by our simulation and displayed
in figure~(\ref{fig:instabilite}). The largest peak of the blue
curve corresponds exactly to the region in the flow where $\beta-U''$
is negative.

\begin{figure}
\includegraphics[scale=0.2]{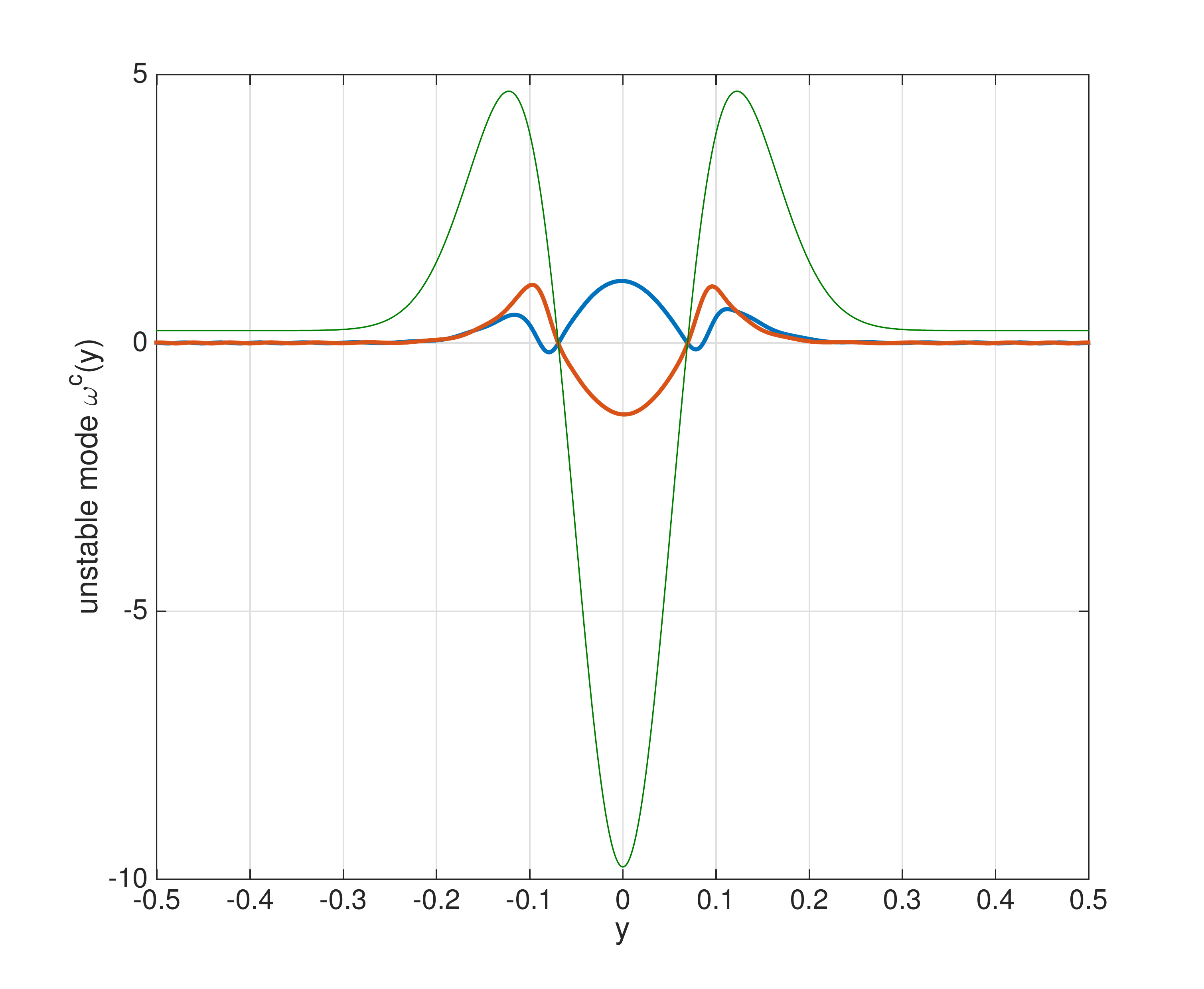}
\includegraphics[scale=0.2]{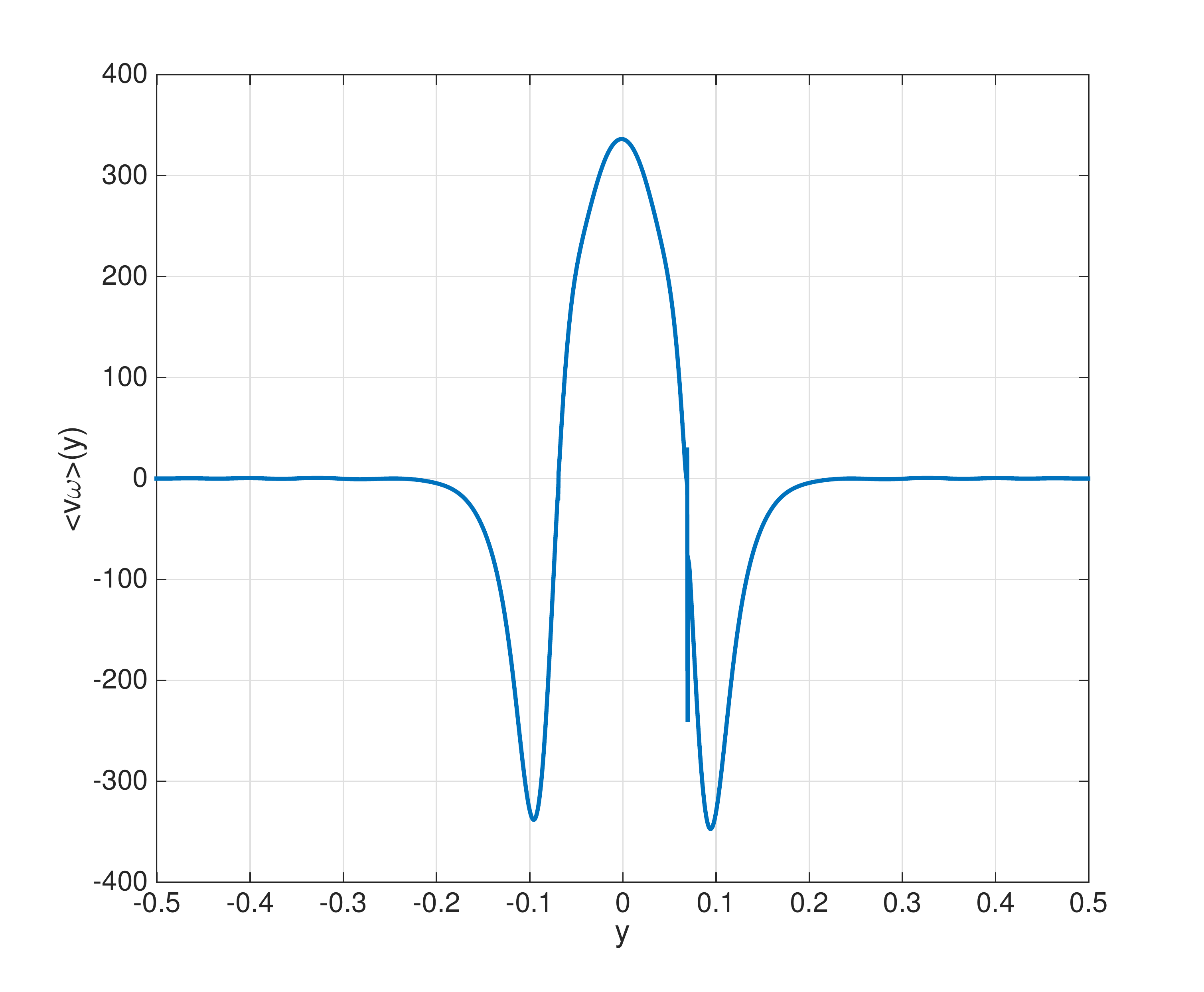}

\protect\caption{Unstable mode at the westward jet edge.
Left: real part (blue curve) and imaginary part (red curve) of the
unstable mode $\omega^{c}$. The value of $\eta$ is $0.05$. The
thiner curve (green) displays $\beta-U''(y)$, the Rayleigh-Kuo criterion
is violated in the vincinity of zero. Right: The tensor $\frac{|\omega|^{2}-1}{U''-\beta}$
that contributes to the Reynolds stress divergence obtained for T=30.
The artefact comes from the fact that $U''-\beta$ vanishes and that
we use a finite discretisation in $y$. The effect of this tensor
is to reequilibrate the velocity $U$ to satisfy $\beta-U''>0$. The
value of $c$ is $c=-2.02+1.04i~e-2$\label{fig:tenseur instable}}
\end{figure}

To obtain the mode $\omega^{c}(y)$, we simply
look at the convergence of $\omega(y,t)e^{ikct}$. The real and imaginary
parts of the unstable mode $\omega^{c}(y)$ are displayed in the left panel of figure~(\ref{fig:tenseur instable}) in respectively blue and
red. The curve $\beta-U''$ has been superimposed in yellow. We see
again that the unstable mode is vanishing at points where $\beta-U''=0$
and that the mode is larger in the region where $\beta-U''<0$. In
the right panel of figure~(\ref{fig:tenseur instable}) we display
the Reynolds stress divergence $\left\langle v\omega\right\rangle =\frac{|\omega|^{2}-1}{U''-\beta}$
obtained from equation (\ref{eq:inertial}) with one single Fourier
component. As can be checked directly in figure~(\ref{fig:tenseur instable})
right, the effect of the instability is exactly the opposite as the one in
figure~(\ref{fig:Reynold for parabola}). The Reynolds stress divergence
is positive in the region where the Rayleigh-Kuo criterion is violated,
and thus the tensor in figure~(\ref{fig:tenseur instable}) reequilibrates
the profile $U$ and damps the perturbation~$-\eta e^{-\frac{y^{2}}{\sigma^{2}}}$.\\

Let us summarize the results of the present section. We have first investigated
the behavior of the Reynolds stress divergence for a parabolic profile,
because  numerical simulations show that the mean velocity
profile is almost parabolic for westward jets. The Reynolds stress
divergence is the tensor that forces the mean flow according to equation
(\ref{eq:meanvelocity}). Even if we cannot always compute exactly
this tensor, we can study its sign and its qualitative properties
to see whether it damps the flow or not. We first did the
assumption that there is no hydrodynamic instability in the flow. This assumption leads to a contradiction for the parabolic profile
because the Reynolds stress divergence distorts the parabolic profile
at $y=0$ as shown in figure~(\ref{fig:Reynold for parabola}).
Thus, we conclude that another mechanism takes place to equilibrate
the parabolic profile. When we consider a small violation of the Rayleigh-Kuo
criterion near $y=0$, we see numerically the growth of an instability
that opposes exactly to the distortion of the parabolic profile where
the Rayleigh-Kuo criterion is violated. Those results are qualitative,
we did not compute the equilibrium velocity profile. But it
is consistent to assume that the equilibration mechanism is a kind
of barotropic adjustment of the mean flow: an instability develops
as soon as $\beta-U''$ changes sign. The flow has to adjust itself
such that the instability is not too large, i.e close to a parabolic
profile with $U''\sim\beta$, and such that the instability can be
damped by linear friction.

\section{Conclusion and perpectives}

The stochastic barotropic $\beta$ plane model is the simplest model in the hierarchy of models aiming at understanding jet formation in atmosphere dynamics. The precise structure of jets in this simple and fundamental model is still not really understood beyond qualitative description and orders of magnitude estimates, although thousands of papers have been written on the subject. In this paper we have proposed three main contributions to the theoretical understanding of these zonal jets to make progresses in this direction. Our analytical results are valid when assuming both the inertial and the small scale forcing limits. The inertial limit is valid when the timescales related to the inviscid dynamics (perfect transport, shearing and mixing, Rossby waves, and so on) are much smaller than the timescale for spin up and spin down (related to forcing and dissipation). In the the stochastic barotropic model, this is quantified by a small value of the nondimensional parameter $\alpha = L\sqrt{r^3/\epsilon}$ . The limit of small scale forces is relevant when the typical scale for the forcing, $1/K$, is much smaller than the typical jet width, of the order of the Rhines scale. Those two limits are relavant for instance for the largest zonal jets of Jupiter.

With these two limits, the interaction between the large scale zonal jets and the small scale turbulence becomes local in physical space. The energy is transferred directly from its injection scale to the zonal jet through the direct interaction between the jet and turbulence. Our first contribution has been to justify the local formula  $\left\langle uv\right\rangle =\frac{\epsilon}{U'}$. This formula could have been obtained directly taking the limit of small friction in the formula of \citep{srinivasan2014reynolds}, or by neglecting the nonlinear and pressure terms in the energy balance as done by \citep{laurie2014universal} for the case of 2D turbulence. Our justification is based on the double limit discussed in the previous paragraph. The mathematical difficulty resides in considering the inertial limit before the limit of small scale forces. This order of the limits is necessary if one wants to deal with situations for which the dissipation mechanism is much smaller than the inertial one at the forcing scale, which is the case for most geophysical turbulent flows. 

Because jets have non-monotonic velocity profiles, the asymptotic expansion and the formula $\left\langle uv\right\rangle =\frac{\epsilon}{U'}$ break down at the jet edges where $U'=0$. The first naive computation of the velocity profile in the limit $K\rightarrow\infty$ leads to a divergence of the velocity at the extremum, and confirms that the dynamics may lead to several velocity sign reversal. For the asymptotic expansion to be valid,
the parameter $\frac{KU'}{U''}$ has to be large.  At the eastward jet edges, we have established that 
the velocity profile is regularized by a cusp at a typical scale of $1/K$. In the inertial limit $\alpha\rightarrow0$, we have derived a system of equations that describes the cusp velocity profile. The resulting shape depends on the forcing spectrum. Nevertheless,
the mechanism of depletion of vorticity at the stationary streamlines leads to an interesting relation between the curvature of the cusp and
the maximal velocity (\ref{eq:curvature}), $U(y_{cr})U''(y_{cr})=-\epsilon K^{2}/r$, which does not depend on the force spectrum but just on the energy injection rate $\epsilon$. 

As observed in previous numerical studies of the barotropic model, the westward jet edges have a curvature $U''$ of order $\beta$ in the inertial limit. Based on observations in numerical studies, the mechanism of barotropic adjustment has been discussed for this selection of the jet curvature (see for instance \citep{constantinou2012emergence}). In the present work we have for the first time derived and analyzed the equations that describe the westward jet edges in the inertial limit. This theoretical analysis confirms the mechanism of barotropic adjustment. Bellow the stability threshold, the Reynolds stress divergence forces the mean flow to grow. But as soon as the mean flow has a curvature $U''>\beta$, an hydrodynamic instability opposes the growth of the velocity profile. Therefore, the parabolic profile is stabilized with a curvature fluctuating close to 
$\beta$. The flow remains close to marginal stability.

Our work gives an overall picture of the equilibration mechanism and the stationary velocity profile of barotropic zonal jets. Our work can be considered as a theoretical derivation that the jet velocity profile is close to the "PV staircase" in the inertial and small scale forcing limit. The "PV staircase" idea is closely related to the qualitative ideas of homogenization of potential vorticity first proposed by Rhines and Young, and can be justified qualitatively by equilibrium statistical mechanics. However those qualitative ideas do not allow for clear predictions. In the context of barotropic jets and Jupiter jets, The "PV staircase" empirical evidence or the "PV staircase" assumption were discussed thoroughly by \citep{Dritschel_McIntyre_2008JAtS}. For instance, \citep{Dritschel_McIntyre_2008JAtS} showed that the "PV staircase" assumption allows to derive straightforwardly the number and the size of jets in a flow configuration. One should bear in mind that the "stairs" are an idealization of the real profile: the discontinuity in the staircase profile corresponds to the cusp at the eastward extremum. It has thus a finite width typically given by $\frac{1}{K}$, where K is the typical wavevector of small-scale energy injection. Besides, the "stairs" are not perfectly flat, because the curvature of the jet is close to $\beta$ at the westward extremum, but not between the extrema. In the monotonic region between the extrema, the mean velocity profile is described by equation (3.9). The "PV staircase" profile can thus be seen as a very good approximation, our theoretical approach gives a more precise mathematical description of the actual profile which is valid within the asymptotic regime of inertial and small scale forcing limit. Moreover, a given flow can sustain different numbers of jets \citep{BakasIoannou2013SSST,constantinou2012emergence} for the same value of the parameter. This observation cannot be predicted neither from the "PV staircase" approximation, nor from the results we presented in this work. Complementing this work results in order to determine the correct number and spacing between those jets is a challenging problem that might be addressed in the future.

Westward jets on Jupiter display a parabolic profile, but with a curvature $U''$ clearly larger than $\beta$. This is one reason why the barotropic model is not sufficient to describe Jupiter's atmosphere. We believe our analysis could be extended to more refined models, for example a two-layer model. The generalization of our analytical results for a two-layer quasi-geostrophic model would be a very interesting extension of this work. Another natural extension would be the study of rare transitions between states with a different number of jets, within the theoretical framework discussed in this paper.

%
\acknowledgments
We thank P. Ioannou for interesting discussion during the preliminary stage of this work.

The research leading to these results has received funding from the European Research Council under the European Union's seventh Framework Program (FP7/2007-2013 Grant Agreement No. 616811).

\appendix

\section{The Reynolds stress divergence in the inertial limit\label{sec:The-Reynold's-stress}}

The aim of this section is to give the proof of formula (\ref{eq:inertial})
and (\ref{eq:inertialmodes}). We have to compute 
\begin{equation}
2\alpha\left\langle |\omega|^{2}\right\rangle =2\alpha\int_{-\infty}^{0}{\rm d}t~e^{2\alpha t}\left|e^{tL_{k}}[c_{l}]\right|^{2},\label{eq:formuleannexeA}
\end{equation}
where $e^{tL_{k}}[c_{l}]:=\omega_{d}$ is the solution to the deterministic
equation
\begin{eqnarray}
\partial_{t}\omega_{d}+ikU\omega_{d}+ik(\beta-U'')\psi_{d} & = & 0\label{eq:perturbationannexeA}\\
\left(\partial_{y}^{2}-k^{2}\right)\psi_{d} & = & \omega_{d}\nonumber 
\end{eqnarray}
with initial condition $c_{l}(y)=e^{ily}$. We will first assume there
are no neutral modes solutions of (\ref{eq:perturbationannexeA}).
First, we do the change of timescale $2\alpha t\rightarrow t$ in
the integral of (\ref{eq:formuleannexeA}). It gives us
\[
2\alpha\left\langle |\omega|^{2}\right\rangle =\int_{-\infty}^{0}{\rm d}t~e^{t}\left|e^{\frac{t}{2\alpha}L_{k}}[c_{l}]\right|^{2}.
\]
When $\alpha$ goes to zero, the term $e^{\frac{t}{2\alpha}L_{k}}[c_{l}]$
is the long time limit of the solution of (\ref{eq:perturbationannexeA}).
We use the nontrivial result for the case of non monotonous flows,
of \citep{Bouchet_Morita_2010PhyD} already mentioned, that there exists
a function $\omega_{d}^{\infty}(y)$ such that $\omega_{d}(y,t)\underset{t\rightarrow\infty}{\sim}\omega_{d}^{\infty}(y)e^{-ikUt}$
when there are no neutral modes. Hence $\left|e^{\frac{t}{2\alpha}L_{k}}[c_{l}]\right|\rightarrow|\omega_{d}^{\infty}(y)|$
, and the presence of the exponential in the integral ensures the
convergence of the whole. This proves that without neutral modes 
\[
2\alpha\left\langle |\omega|^{2}\right\rangle \underset{\alpha\rightarrow0}{\longrightarrow}|\omega_{d}^{\infty}|^{2}.
\]

The second case, with neutral modes, is a bit more subtle. The result
of \citep{Bouchet_Morita_2010PhyD} relies on a Laplace transform of
$\omega_{d}$ denoted $\hat{\omega}_{d}$. To do the inverse Laplace
transform, one has to know where the singularities of \textbf{$\hat{\omega_{d}}$}
are. The presence of modes in the equation is exactly equivalent to
the presence of poles of order 1 in the complex plane for $\hat{\omega}_{d}$.
For unstable modes, these poles have an imaginary part, whereas for
neutral modes, they are located on the real axis. We also assume in
our calculation that there are no instabilities, which means that
all singularities of $\hat{\omega}_{d}$ are on the real axis. Some
of these singularities are outside the range of $U$ (outside of {[}$U_{min},U_{max}${]})
and are isolated, they correspond to neutral modes or ``modified
Rossby waves''. But there is also a continuum of singularities all
along the range of U. The integration around the isolated singularities
will give the contribution of neutral modes, and it is of the form
$\underset{a}{\sum}\omega^{a}(y)e^{ikc_{a}t}$ where $a$ is the mode
index, $c_{a}$ is the mode frequency, and the $\omega^{a}(y)$ are
the projections of the initial condition $c_{l}$ on the modes $\zeta^{a}(y)$.
The projections are defined with the natural scalar product induced
by the pseudomomentum conservation law, that is $\ll\omega_{1}^{*}\omega_{2}\gg=\int\frac{\omega_{1}^{*}\omega_{2}}{U''-\beta}{\rm d}y$.
For this particular scalar product, the operator $\omega\rightarrow U\omega+(\beta-U'')\psi$
is self-adjoint, and this implies that its eigenvectors are orthogonal
with respect to this scalar product. We substract the contribution
of the modes $\zeta^{a}(y)$ from the initial condition $c_{l}$, that is, we use $c_l-\underset{a}{\sum}\omega^{a}$ as initial condition in (\ref{eq:perturbationannexeA}).  We are left with the continuum part of the
singularities and the result of \citep{Bouchet_Morita_2010PhyD}
holds. As a consequence, there exists a function $\tilde{\omega}_{d}^{\infty}(y)$
such that the remaining part of the solution behaves at infinity like
$\tilde{\omega}_{d}^{\infty}(y)e^{-ikU(y)t}.$ We eventually find
that for long time, the solution of the deterministic equation behaves
like
\[
\omega_{d}(y,t)\underset{t\rightarrow\infty}{\sim}\underset{a}{\sum}\omega^{a}(y)e^{-ikc_{a}t}+\tilde{\omega}_{d}^{\infty}(y)e^{-ikU(y)t}.
\]

When we inject this result in the expression of $\left|e^{\frac{t}{\alpha}L_{k}}[e_{l}]\right|^{2}$
we get three different terms.
\begin{enumerate}
\item Terms coming from the mode-mode contribution of the form $\underset{a}{\sum}|\omega^{a}(y)|^{2}$.
The time integration is then trivial.
\item The term coming from the continuum gives us immediately the contribution
$|\tilde{\omega}_{d}^{\infty}(y)|^{2}$.
\item What happens for terms of the form $\omega^{a*}\tilde{\omega}^{\infty}e^{-ik(U-c_{a})\frac{t}{\alpha}}$
and $\omega^{a*}\omega^{b}e^{-ik(c_{b}-c_{a})\frac{t}{\alpha}}$ ?
The frequencies $\frac{1}{\alpha}k(U-c_{a})$ and $\frac{1}{\alpha}k(c_{a}-c_{b})$
grow to infinity as $\alpha$ vanishes. We have an oscillating integral
with frequency growing to infinity. It is a well known result that
such an integral asymptotically decays. The cross terms gives no contributions. 
\end{enumerate}
We have then proved the desired result that 
\[
2\alpha\left\langle |\omega|^{2}\right\rangle \underset{\alpha\rightarrow0}{\longrightarrow}\underset{a}{\sum}|\omega^{a}|^{2}+|\tilde{\omega}_{d}^{\infty}|^{2}.
\]

\section{Computation of the Reynolds stress in the inertial and small scale forcing regime \label{sec:Equivalence-of-the_limits}}

In this appendix we prove that 
\[
\mathcal{R}e\left\langle v_{\theta}^{*}\omega_{\theta}\right\rangle \underset{K\rightarrow\infty\alpha\rightarrow0}{\rightarrow}\frac{U''}{U'^{2}}\label{eq:resultat-annexeB1}.
\]

In section \ref{sec:Inertial-small-scale}, we found the expression
\[
\mathcal{R}e\left\langle v_{\theta}^{*}\omega_{\theta}\right\rangle \underset{K\rightarrow\infty}{\sim}-\frac{\hat{C}_{k,l}}{2k^{2}}\mathcal{P}\left\{ \int{\rm d}Ye^{-|Y|}\frac{\cos(Y\tan\theta)}{U\left(y-\frac{Y}{k}\right)-U(y)}\right\} .
\]
This expression has no meaning for $y_{c}$ such that $U'(y_{c})=0$
because we have a quadratic singularity in the integral. We therefore
assume that $U'(y)$ does not vanish.
With this assumption
we get
{\small 
\begin{eqnarray*}
 \mathcal{P}\left\{ \int{\rm d}Ye^{-|Y|}\frac{\cos(Y\tan\theta)}{U\left(y-\frac{Y}{k}\right)-U(y)}\right\}
 & =  \underset{\eta\rightarrow0}{\lim}\int_{-\infty}^{-\eta}{\rm d}Ye^{-|Y|}\frac{\cos(Y\tan\theta)}{U\left(y-\frac{Y}{k}\right)-U(y)}+\int_{\eta}^{+\infty}{\rm d}Ye^{-|Y|}\frac{\cos(Y\tan\theta)}{U\left(y-\frac{Y}{k}\right)-U(y)}\\
 & =  \int_{0}^{+\infty}{\rm d}Ye^{-|Y|}\cos(Y\tan\theta)\left\{ \frac{1}{U\left(y-\frac{Y}{k}\right)-U(y)}+\frac{1}{U\left(y+\frac{Y}{k}\right)-U(y)}\right\} \\
 & =  \int_{0}^{+\infty}{\rm d}Ye^{-|Y|}\cos(Y\tan\theta)\left\{ \frac{U\left(y-\frac{Y}{k}\right)+U\left(y+\frac{Y}{k}\right)-2U(y)}{\left(U\left(y-\frac{Y}{k}\right)-U(y)\right)\left(U\left(y+\frac{Y}{k}\right)-U(y)\right)}\right\} \\
 & \underset{k\rightarrow\infty}{\rightarrow}  \frac{U''}{U'^{2}}\int_{0}^{+\infty}{\rm d}Ye^{-|Y|}\cos(Y\tan\theta)=\frac{U''}{U'^{2}}\frac{1}{1+\tan^{2}\theta}.
\end{eqnarray*}}
We have used the relations 
$$
U\left(y-\frac{Y}{k}\right)+U\left(y+\frac{Y}{k}\right)-2U(y)\sim U''\frac{Y^{2}}{k^{2}}
$$
and 
$$
\left(U\left(y-\frac{Y}{k}\right)-U(y)\right)\left(U\left(y+\frac{Y}{k}\right)-U(y)\right)\sim U'^{2}\frac{Y^{2}}{k^{2}}.
$$

We have thus
\[
\mathcal{R}e\left\langle v_{\theta}^{*}\omega_{\theta}\right\rangle \underset{K\rightarrow\infty}{\sim}-\frac{\hat{C}_{k,l}}{2k^{2}}\frac{U''}{U'^{2}}\frac{1}{1+\tan^{2}\theta}=-\frac{\hat{C}_{k,l}}{2K^{2}}\frac{U''}{U'^{2}}\label{eq:resultat-annexeB}.
\]
Finally, we use the power input relation $\frac{1}{2}\iint{\rm d}k'{\rm d}l'\frac{\hat{C}_{k',l'}}{K'^{2}}=1$.
When we integrate relation (\ref{eq:resultat-annexeB}) over the whole spectrum, we get the desired result (\ref{eq:resultat-annexeB1}).

\section{Modified Rossby waves\label{sec:Modified-Rossby-waves}}

In this appendix, we discuss the neutral modes of parabolic jets $U(y)=\gamma\frac{y^{2}}{2}$. 
We first note that for any parabolic jet, $U''-\beta$ does not change
sign. Hence the Rayleigh-Kuo criteria is satisfied and the jet has
no unstable modes. 

\begin{figure}
\begin{centering}
\includegraphics[scale=0.25]{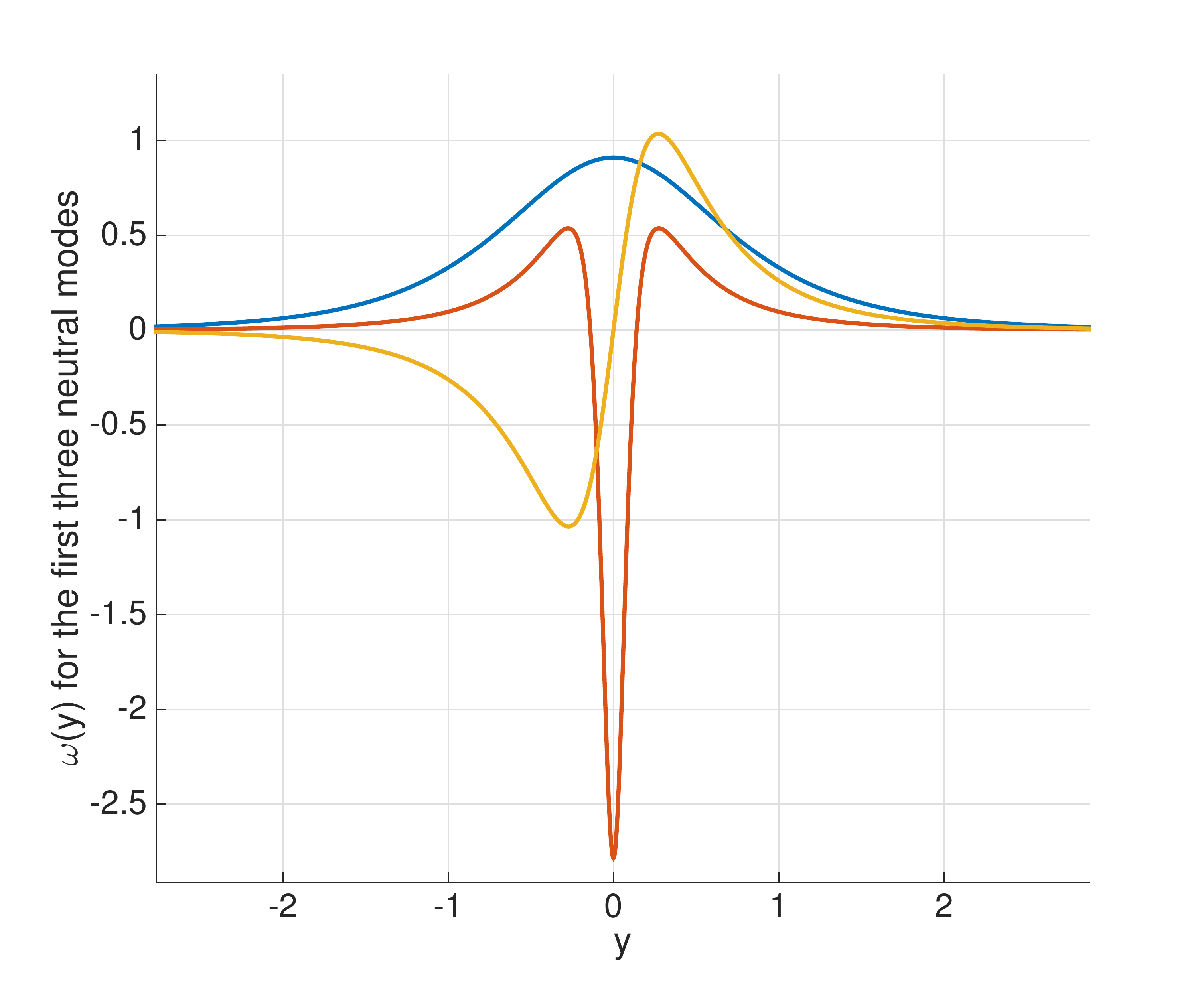}
\par\end{centering}
\caption{The first three neutral modes (modified Rossby waves) obtained for
$\mu=-1$. Even modes and odd modes alternate. The first mode vanishes
only at infinity, and each new mode has an additional zero.}
\label{fig:modes}
\end{figure}

For a mean velocity $U(y)=\gamma\frac{y^{2}}{2}$, we are looking
for solutions of (\ref{eq:deterministic}) of the form $\omega_{k}(y)e^{-ikct}$.
The equation writes
\[
-ikc\omega_{k}+ik\gamma\frac{y^{2}}{2}\omega_{k}+ik(\beta-\gamma)H_{k}*\omega_{k}=0.
\]
We do the Fourier transform $\hat{\omega}_{k}(l):=\int{\rm d}y\omega_{k}(y)e^{-ily}$
to obtain 
\begin{equation}
-\frac{1}{2}\frac{{\rm d}^{2}}{{\rm d}l^{2}}\hat{\omega}_{k}+\frac{\mu}{k^{2}+l^{2}}\hat{\omega}_{k}=\frac{c}{\gamma}\hat{\omega}_{k},\label{eq:schrodinger}
\end{equation}
where $\mu:=1-\frac{\beta}{\gamma}$. We recognize the eigenvalue
problem for a Schr{\"o}dinger operator for a particle with  potential $\frac{\mu}{k^{2}+l^{2}}$,
a result already obtained by Brunet \citep{brunet1990dynamique}. Using
classical results for this type of 1D Schr{\"o}dinger equation, we can
immediately conclude that
\begin{itemize}
\item There exists a solution $\hat{\omega}_{k}$ in $L_{2}$ iff $\mu<0$
(attractive potential). In that case, the corresponding eigenvalue
is negative which implies that $\frac{c}{\gamma}<0.$ The condition
$\mu<0$ imposes already $\gamma>0$, so the phase velocity of Rossbywave
is $c<0$ and the wave is outside of the continuous spectrum. We find
for this particular configuration the classical result that Rossby
waves propagate with $c<U_{min}$.
\item The number of modes $n\left(\left|\mu\right|\right)$ increases with
$\left|\mu\right|$, the depth of the potential well. Modes organize
into continuous families $\left\{ \Omega_{i}\left(\left|\mu\right|\right)\right\} _{1\leq i\leq n\left(\left|\mu\right|\right)}$,
with energies $E_{i}\left(\left|\mu\right|\right)=\frac{c}{\gamma}$,
when $\left|\mu\right|$ is changed. $E_{i}\left(\left|\mu\right|\right)$
are decreasing functions of $\left|\mu\right|$. The families $\Omega_{i}$
are alternatively even and odd functions with a number of nodes that
increases with $i$. A new set of modes appears for critical values
$\mu_{i}$. For $\mu=\mu_{i}$, the mode of the new family has a zero
energy $E_{i}\left(\left|\mu_{i}\right|\right)=0$.
\end{itemize}
To compute the eigenfunction of (\ref{eq:schrodinger}), for a given
$\mu$ we use a bisection algorithm. We divide the interval $[\mu,0]$
in sufficiently small intervals $[\tau_{i},\tau_{i+1}]$ and compute
the solution of (\ref{eq:schrodinger}) with $\frac{c}{\gamma}=\tau_{i}$.
The solution diverges like an exponential at infinity, and when this
divergence changes sign between $\tau_{i}$ and $\tau_{i+1}$, it
means that we have an eigenfunction in the interval, and we iterate
the algorithm until $\tau_{i+1}-\tau_{i}$ is small enough. This way,
we obtain the Fourier transform of a mode, we just have to inverse
the Fourier transform to get the mode in real space. Then we project
$e^{ily}$ on these modes with the standard scalar product on $L_{2}$.
Figure~(\ref{fig:modes}) displays the 3 first eigenfunctions obtained
with $\mu=-1$.

\bibliographystyle{jfm}
\bibliography{All-2016-05,Fbouchet_2016_05}

\end{document}